# Efficient and Secure Mobile Cloud Networking

## Jacques Bou Abdo

# Pierre and Marie Curie University

Doctoral School of Computer science, Communication and Electronics

Submitted in partial fulfillment of the requirements for the degree of
**DOCTOR OF SCIENCE of the Pierre and Marie Curie University**

Specialization
**COMPUTER SCIENCE AND NETWORKING**

## Efficient and Secure Mobile Cloud Networking

# Jacques Bou Abdo

Defended publicly on 18 Dec. 2014.

**Committee in charge:**

| | | |
|---|---|---|
| **Francine Krief** | **Reviewer** | **Professor at Bordeaux University** |
| **Nathalie Mitton** | **Reviewer** | **Head of research group at Inria** |
| **Pascal Urien** | **Member** | **Professor at Telecom ParisTech** |
| **Stefano Secci** | **Member** | **Assoc. Professor. at UPMC - Sorbonne University** |
| **Prosper Chemouil** | **Member** | **Expert Program Leader at Orange Labs** |
| **Jacques Demerjian** | **Co-advisor** | **Assoc. Professor at Lebanese University** |
| **Kablan Barbar** | **Co-advisor** | **Professor at Lebanese University** |
| **Guy Pujolle** | **Advisor** | **Professor at UPMC - Sorbonne University** |
| **Hakima Chaouchi** | **Advisor** | **Professor at Telecom SudParis** |

# Abstract

Mobile cloud computing is a very strong candidate for the title "Next Generation Network" which empowers mobile users with extended mobility, service continuity and superior performance. Users can expect to execute their jobs faster, with lower battery consumption and affordable prices; however this is not always the case. Various mobile applications have been developed to take advantage of this new technology, but each application has its own requirements. Several mobile cloud architectures have been proposed but none was suitable for all mobile applications which resulted in lower customer satisfaction. In addition to that, the absence of a valid business model to motivate investors hindered its deployment on production scale.

This dissertation proposes a new mobile cloud architecture which positions the mobile operator at the core of this technology equipped with a revenue-making business model. This architecture, named OCMCA (Operator Centric Mobile Cloud Architecture), connects the user from one side and the Cloud Service Provider (CSP) from the other and hosts a cloud within its network. The OCMCA/user connection can utilize multicast channels leading to a much cheaper service for the users and more revenues, lower congestion and rejection rates for the operator. The OCMCA/CSP connection is based on federation, thus a user who has been registered with any CSP, can request her environment to be offloaded to the mobile operator's hosted cloud in order to receive all OCMCA's services and benefits.

The contributions of this dissertation are multifold. First, we propose OCMCA and prove that it has superior performance on all other mobile cloud architectures. The business model of this architecture focuses on user's subscription freedom, i.e. the user can be subscribed with any cloud provider and still be able to connect through this architecture to her environment with the help of offloading and federation. Since OCMCA offers services to mobile users who should be authenticated first with the mobile operator before authenticating with the CSP to gain access to her environment and registered services, we propose a robust authentication and single-sign-on protocol, named EC-AKA3 (Ensured Confidentiality Authentication and Key Agreement protocol version 3), capable of performing both authentications in parallel. This protocol achieves faster response than currently existing mechanisms and achieves secure and private authentication at both NAS (Non-Access Stratum) and application layers.

Second, we study privacy problems in various mobile cloud applications and show that privacy preserving mechanisms implemented at existing mobile cloud architectures fail to offer satisfactory levels. We also show that OCMCA can offer higher privacy levels for the discussed applications and can be extended to become a lawful interception interface.

Third, we prove, using a mathematical model, that our proposition to use federation is financially feasible. We also prove that unmonitored federation might result in catastrophic impact on performance, delay and network congestion. To solve this problem we propose a new cloud federation manager called BBCCFM (Broker-Based Cross-Cloud Federation Manager), to be used by OCMCA. This manager facilitates the selection of the federation offers while monitoring it to prevent the shown hazards. BBCCFM results in lower delay, traffic and cost by consolidating requests at a centralized node (Broker) and forming economy of scale. BBCCFM has a comparable availability to other distributed mechanisms and is compliant with the recommendations of "Cloud Security Alliance".

# Key Words:

Mobile Cloud Computing, business model, security, privacy, crowdsourced location based services, cloud federation, EPS, AKA.

# Table of Contents

# CHAPTER 1 INTRODUCTION

## 1.1. Introduction

Cloud computing is a concept reforming IT industry and consequently changing how companies look into IT infrastructure. Reduced CAPEX (Capital Expenditures) and pay-as-you-go mode is best suiting startups and SMEs (Small and Medium Enterprises) to outsource their computation resources [1] [2] and focus their capital on creating a competitive edge against other products and services. Even though many companies (especially major ones) are reluctant to migrate their data to the cloud [1], non-corporate end users are easily adopting this new trend. Weak privacy and business models are two main aspects delaying wide adoption of computation outsourcing [3]and hindering an investment boom until having all these concerns tackled.

With its mobility, reduced latency and increased bandwidth, mobile networks are becoming the network of choice for many corporate and non-corporate users to connect to Internet. Mobile devices are increasingly popular and currently constituting 40% of Internet accessing devices [4].Cloud-based mobile applications are expected to reach 9.5$ billion by end of 2014 [4]. Cloud and mobile computing are performing a successful and very promising tag team which is estimated to dominate the storage and processing traffic over the Internet.

Mobile cloud computing is a very strong candidate for the title "The Next Generation Network" which empowers mobile users with extended mobility, service continuity and superior performance. Its main concept lays in offloading a job from a resource-limited mobile device to be executed at the cloud and then receiving the result once done. Users can expect to execute their jobs faster, with lower battery consumption and affordable prices; however this is not always the case. Various mobile applications have been developed to take advantage of this new technology, but each application has its own requirements. Several mobile cloud architectures have been proposed in literature but none was suitable for all mobile cloud applications and this leads to low customer satisfaction for all the customers using the applications not suited by the implemented architecture.

Having a mobile cloud architecture suitable for all applications is simply a must, but it is not enough. A valid business model that motivates investors is a vital factor in making any technology reach deployment phase. This fact emphasizes the need for a technology capable of satisfying the requirements of both major players (users and investors) who control the dynamics of market strategy.

This thesis targets at proposing competitive mobile cloud architecture based on an interesting business model which motivates investors, especially mobile operators. This architecture should be future-proof by being compatible with horizontal federation, the destiny of maturing industries. This thesis also targets at proposing a multi-layer private authentication and single-sign-on protocol in addition to various privacy applications on the proposed mobile cloud architecture. In the following, the main objectives, challenges and our approaches for achieving this goal are described.

## 1.2. Objectives and Workflow

Mobile cloud computing is a very promising technology that provides its users a wide range of services and its operators elevated user traffic. This trend is putting the infrastructure providers (mobile operators) face to face with critical problems such as: demanding QoS (Quality of Service), investment optimization, user privacy, scalability etc. We are able to tackle these problems in four interdependent and incremental objectives:

- **Objective 1: Resource optimizing mobile cloud architecture**. We proposed an architecture (titled OCMCA: Operator Centric Mobile Cloud Architecture) taking into consideration its ability to offer very competitive QoS which satisfies users' requirements. The QoS parameters used in evaluating the architecture's performance are specifically selected to represent the users' interest. This architecture is able to achieve high profits, penetration rates and investment optimization. This objective is met in chapter 2. Reference papers are:

- • J. Bou Abdo, J. Demerjian, H. Chaouchi, K. Barbar and G. Pujolle, "Operator Centric Mobile Cloud Architecture", IEEE Wireless Communications and Networking Conference (IEEE WCNC 2014).

- **Extended privacy offered by the mobile cloud architecture**. The proposed architecture should offer elevated privacy level for its users. We divide the provided privacy mechanisms into two categories:

  - **Objective 2: Pre-authentication**: This category contains the AKA (Authentication and Key Agreement) and SSO (Single-Sign-On) mechanisms which will be executed in the absence of a secure context to establish one. This objective is met in chapter 3 using a protocol titled EC-AKA3 (Ensured Confidentiality Authentication and Key Agreement protocol version 3). Reference papers are:

    - • J. Bou Abdo, H. Chaouchi and M. Aoude, "Ensured Confidentiality Authentication and Key Agreement Protocol for EPS". 3rd Symposium on Broadband Networks and Fast Internet, 28-29 May 2012. IEEE.

    - • J. Bou Abdo, H. Chaouchi and J. Demerjian, "Security v/s QoS for LTE Authentication and Key Agreement protocol.", International Journal of Network Security & Its Applications Special Issue on: "Communications Security & Information Assurance". Sept 2012.

    - • J. Bou Abdo, J. Demerjian, K. Ahmad, H. Chaouchi and G. Pujolle, "EPS mutual authentication and Crypt-analyzing SPAKA". International Conference on Computing, Management and Telecommunications (ComManTel 2013), 22-24 Jan 2013. IEEE.

    - • J. Bou Abdo, J. Demerjian, H. Chaouchi and G. Pujolle, "EC-AKA2 a revolutionary AKA protocol". International Conference on Computer Applications Technology (ICCAT 2013), 20-22 Jan 2013. IEEE.

- • J. Bou Abdo, J. Demerjian, H. Chaouchi, K. Barbar and G. Pujolle, "Single-Sign-on in Operator Centric Mobile Cloud Architecture", 17th IEEE Mediterranean Electrotechnical Conference (IEEE Melecon 2014).

- **Objective 3: Post-authentication**: This category contains location-privacy-preserving-mechanisms and identity privacy/investigation dilemma which takes place in the presence of a trust context. This objective is met in chapter 3. Reference papers are:

    - • J. Bou Abdo, J. Demerjian and H. Chaouchi, "Security in Emerging 4G Networks", Next-Generation Wireless Technologies, ISBN 978-1-4471-5164-7, Springer 2013.

    - • J. Bou Abdo, J. Demerjian, H. Chaouchi, K. Barbar and G. Pujolle, "Privacy in Crowdsourcing location based services", 3rd IEEE International Conference on Cloud Networking (IEEE CloudNet).

- **Ensuring scalability through federation.** Future-proof mobile cloud architecture should consider its scalability, especially with the forecasted increase in traffic. Two aspects should be considered to ensure system-wide scalability which are:

    - **Access Network scalability**. If the current 5G research initiative succeeded in achieving its targets, access network scalability can be ensured for undetermined period. This aspect is considered outside the scope of this manuscript.

    - **Objective 4: Computation/Storage scalability**. Federation is considered an important scalability and business continuity factor in cloud computing. A federation establishment mechanism should be proposed to maintain the financial feasibility of this scalability solution. This objective is met in chapter 4. Reference papers are:

        - • J. Bou Abdo, J. Demerjian, H. Chaouchi, K. Barbar and G. Pujolle, "Broker-Based Cross-Cloud Federation Manager", 8[th] International Conference for Internet Technology and Secured Transactions (ICITST-2013). IEEE.

•• J. Bou Abdo, J. Demerjian, H. Chaouchi, K. Barbar and G. Pujolle, "Macro-economy effect on cloud federation", 3rd IEEE International Conference on Cloud Networking (IEEE CloudNet).

•• J. Bou Abdo, J. Demerjian, H. Chaouchi, K. Barbar and G. Pujolle, "Federation means cash ", 3rd International Conference on e-Technologies and Networks for Development (ICeND 2014). IEEE.

•• J. Bou Abdo, J. Demerjian, H. Chaouchi, K. Barbar and G. Pujolle, "Cloud federation? We are not ready yet", 6th International Symposium on Cyberspace Safety and Security (CSS 2014). IEEE.

The objectives and the contribution workflow are presented graphically in figure 1.1.

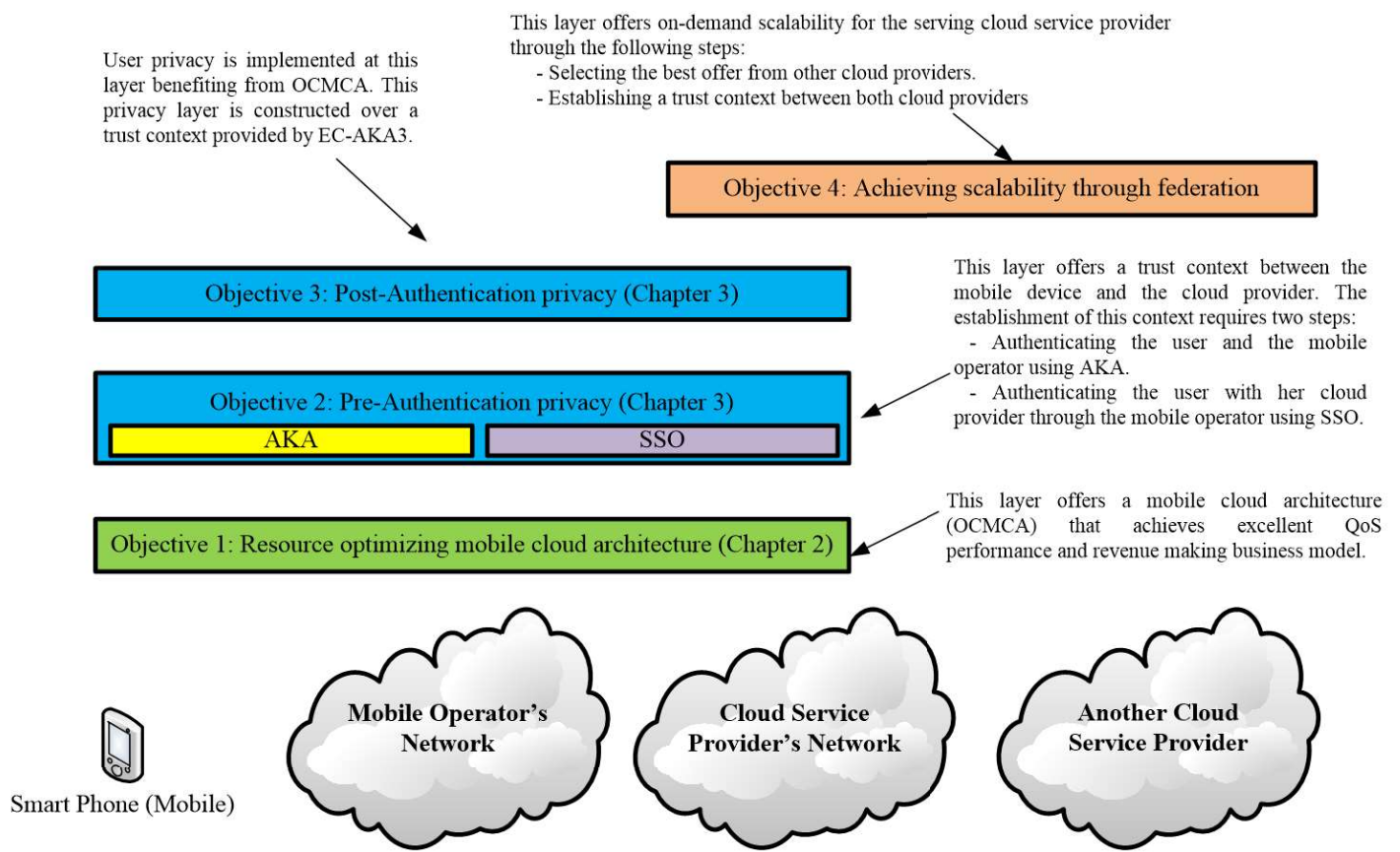

Figure 1.1 Contribution division

# 1.3. Background

## 1.3.1. Cloud computing

Many definitions for cloud computing can be found in literature, each trying to give a complete and specific explanation of what this technology is all about. We start by surveying these definitions and then give our own based on the lessons learnt. Cloud computing is:

- "a model for enabling ubiquitous, convenient, on-demand network access to a shared pool of configurable computing resources that can be rapidly provisioned and released with minimal management effort or service provider interaction." [5].
- "a parallel and distributed computing system consisting of a collection of inter-connected and virtualized computers that are dynamically provisioned and presented as one or more unified computing resources based on Service-Level Agreements (SLA) established through negotiation between the service provider and consumers." [6][7].
- "a large pool of easily usable and accessible virtualized resources (such as hardware, development platforms and/or services). These resources can be dynamically reconfigured to adjust to a variable load (scale), allowing also for an optimum resource utilization. This pool of resources is typically exploited by a pay-per-use model in which guarantees are offered by the Infrastructure Provider by means of customized Service Level Agreements." [7][8].
- "hardware-based service offering compute, network, and storage capacity where: Hardware management is highly abstracted from the buyer, buyers incur infrastructure costs as variable OPEX, and infrastructure capacity is highly elastic." [7][9].

We believe that cloud computing can be better described by two complementary definitions, one conveying the user's perspective and the other conveying the provider's. So we define cloud computing to be:

- Online, infinite-like, easily provisioned and pay-as-you-go resource pool that are bound by SLAs.

- Pool of resource under the management of a centralized entity that is capable of dynamically provisioning it.

The user can offload her job (computation, storage, etc.) to the cloud to decrease the utilization at her local devices and to get additional functionalities she doesn't have locally. The advantages and disadvantages of cloud computing as described in [10] are shown in table 1.1.

Table  1.1 Advantages and disadvantages of cloud computing

| Advantages | Disadvantages |
| --- | --- |
| Lower-Cost Computers for Users | Requires a Constant Internet Connection |
| Improved Compatibility Between Operating Systems | Doesn't Work Well with Low-Speed Connections |
| Lower IT Infrastructure Costs | Features Might Be Limited |
| Fewer Maintenance Issues | Stored Data Might Not Be Secure |
| Lower Software Costs | If the Cloud Loses Your Data, You're Screwed |
| Instant Software Updates | |
| Increased Computing Power | |
| Unlimited Storage Capacity | |
| Increased Data Safety | |
| Improved Performance | |
| Improved Document Format Compatibility | |
| Easier Group Collaboration | |
| Universal Access to Documents | |
| Latest Version Availability | |
| Removes the Tether to Specific Devices | |

The cloud architecture stack is composed of three main layers [11]: SaaS (Software as a Service), PaaS (Platform as a Service) and IaaS (Infrastructure as a Service) as shown in figure 1.2. Additional service layers are continuously proposed to add supporting features, such as PasS[12] and HuaaS[11].

IaaS is the only presented layer, in this introduction, because of its importance to inter-cloud federation which is thoroughly discussed in chapter 4. It is the bottom layer at the cloud stack and nearest to the hardware. It enables on-demand provisioning of servers [7]. IaaS offers two services: VRS (Virtual Resource Set) and PRS (Physical Resource Set) [13]. PRS is hardware dependant and creates an abstraction layer to be used by VRS. VRS is hardware independent and used to monitor virtualized applications. One of the most dominant VRS monitor technologies is called hypervisor or VMM (Virtual Machine Monitor) where the virtualized applications are user virtual machines [7]. VIM (Virtual Infrastructure Manager), regardless of its underlying VMM layer, provides the tools for scheduling and managing VMs across multiple physical layers [14]. VRS, PRS, VIM and VMM are shown in figure 1.2.

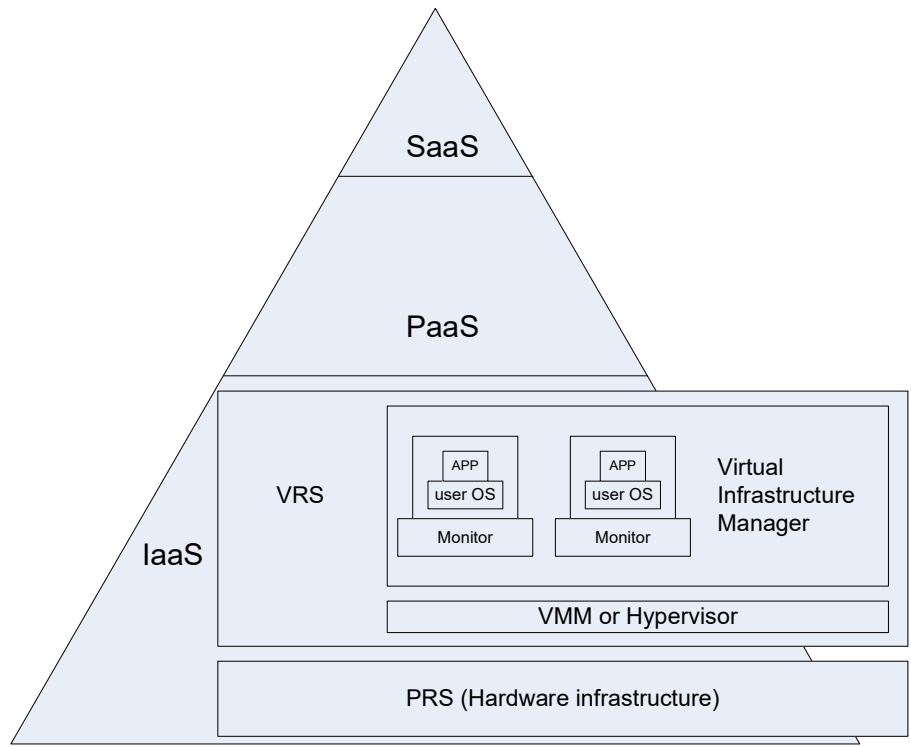

Figure 1.2 Cloud stack (Well-known layers)

Cloud Manager Layer is still needed to completely define the cloud. It provides cloud-like interfaces and higher-level functionalities for authentication, identity management, contextualization and VM disk image management [16]. The cloud manager layer contains also the federation manager [15] which performs the following abstracted operations:

- The first operation is selecting a cloud provider (foreign) having its IdP (Identity Provider) trusted by the client's (the cloud experiencing saturation and interested in federation) IdP, then establishing a secure connection after successful authentication.

- The second operation is transferring the VMs (Virtual Machine) and extending the hypervisor.

### 1.3.2. Mobile cloud computing

Mobile cloud has been considered by [17] to be a collection of mobile devices within the same vicinity all having interest in processing the same data. In this case, the processing cost (battery power and CPU cycles) will be divided on the participating devices and hopefully fulfilling mobile cloud's goals. Mobile cloud is controversial since we are expecting one architecture to be adequate for all mobile applications, always decreasing power consumption in mobile devices and most importantly cheap. No architecture found in literature was able to satisfy the above expectation, and none was standardized.

Mobile cloud computing has been understood differently by the research community and this explains the deep difference in defining this technology and designing its architectures. To give a generalized definition which includes all points of view and architectures, it should be abstract and doesn't specify detailed features. For this reason, we define mobile cloud computing as: "A technology which allows the user to access cloud services through mobile devices".

### 1.3.3.  EPS architecture

EPS (Evolved Packet System) [18] is the $4^{th}$ generation of an extensively used mobile communication system, which has at least 30 years of cumulative experience in large-scale deployments and global interoperability. It is the first all-IP 3GPP release which was designed to be backward compatible with previous releases.

Although EPS offers considerably low latency and high traffic rates, current standardization efforts are attempting to enhance mobile networks' performance to be able to cope with the increasing user demand. These attempts are mainly focusing on RAN (Radio Access Network) with minor modification on the core resulting in future proof services.

In addition to its elevated QoS performance, EPS has the following benefits [19]:

- Reduced cost per bit.
- Increased service provisioning.

- Flexible use of existing and new frequency bands.

- Simplified architecture and open interfaces.

- Reasonable terminal power consumption.

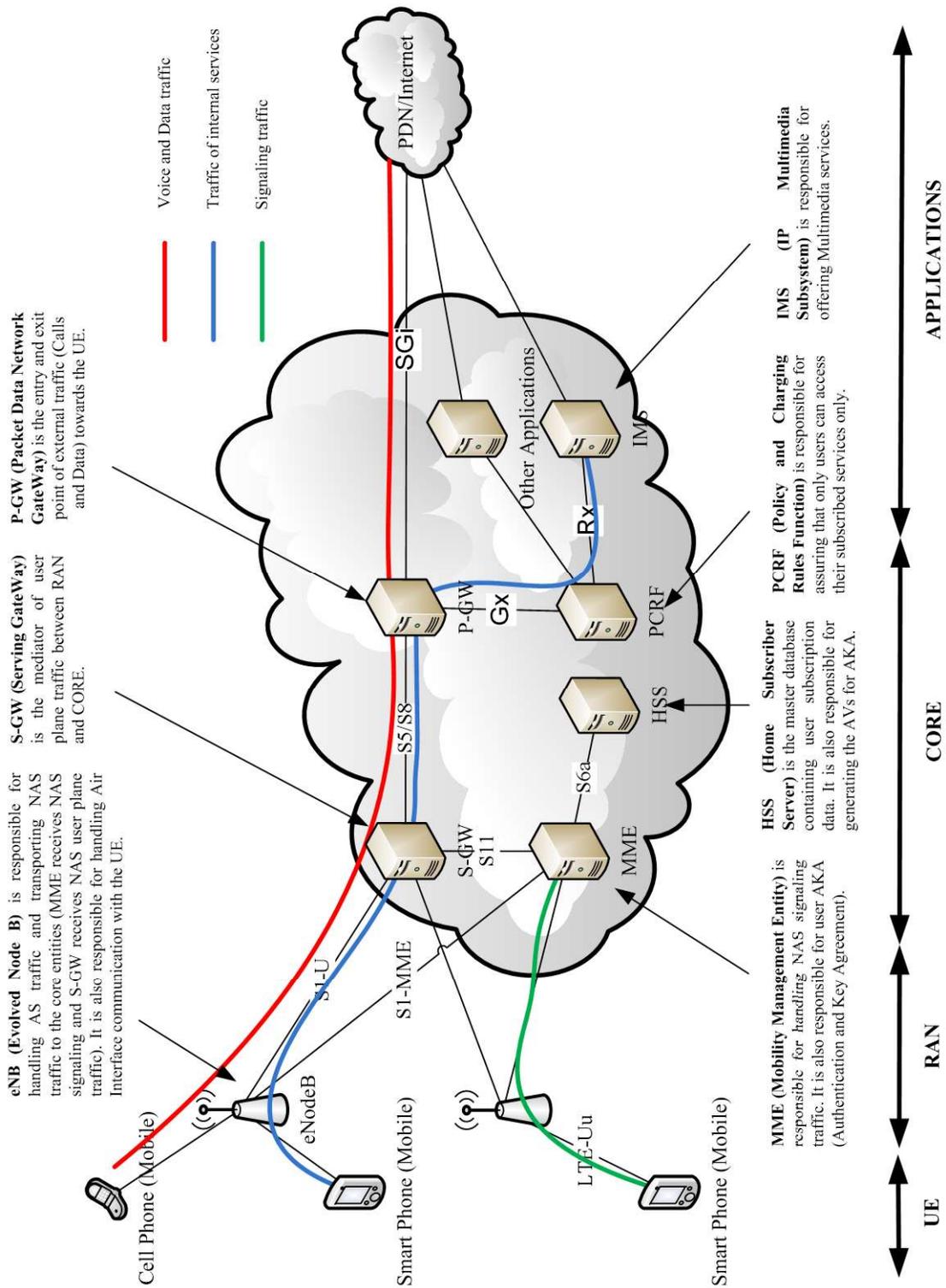

Figure 1.3 EPS architecture

EPS architecture is shown in figure 1.3, where the following entities are presented:

- **UE (User Equipment)**: is the device which allows the user to access the network [20]. Some of its functions are [21]:
  - Contains the authentication information which will be used to authenticate the user during EPS-AKA (Authentication and Key Agreement) as will be shown in the coming section.
  - Supports LTE uplink and downlink air interface.

- **eNB (Evolved Node B)**: is the entity responsible for delivering user and control plane traffic to UE over the air channel. Some of its functions are [21]:
  - IP header compression and ciphering of user data stream.
  - Radio resource management.

- **MME (Mobility Management Entity)**: is the entity responsible for handling NAS (Non-Access Stratum) [22] control plane traffic also known as signaling. Some of its functions are [21]:
  - Tracking Area list management.
  - Authentication and Key Agreement.
  - Lawful Interception of signaling traffic.

- **HSS (Home Subscriber Server)**: is defined in the standard as: "It is the entity containing the subscription-related information to support the network entities actually handling calls/sessions" [20]. It is a very important entity in AKA as will be shown in the coming section. Some of its functions are [21]:
  - Stores the subscriber data.
  - Generates the AVs (Authentication Vector) used in AKA.
  - Used in locating the current position of the user (on MME layer) i.e. finds which MME the user is currently attached to.

- **S-GW (Serving GateWay)**: is RAN's interface with the core network for user plane traffic. Some of its functions are [21]:
  - Lawful Interception [23-25].
  - Packet routing and forwarding [20].
  - Local anchor point for inter-eNB handover.

- **P-GW (PDN-GateWay)**: acts as the mobile network's gateway router toward PDN (Packet Data Network) for both LTE (E-UTRAN) [79] and pre-LTE (GERAN and UTRAN) technologies. Some of its functions are [21]:
  - Per-user based packet filtering.
  - UE IP addresses allocation.

- **PCRF (Policy and Charging Rules Function)**: is the entity responsible for services charging control. It is described in the standard as: "In order to allow for charging control, the information in the PCC (Policy and Charging Control) rule identifies the service data flow and specifies the parameters for charging control. The PCC rule information may depend on subscription data. For the purpose of charging correlation between application level (e.g. IMS) and service data flow level, applicable charging identifiers shall be passed along within the PCC architecture, if such identifiers are available" [26]. Some of its functions are decides how services shall be treated in the PDN gateway [21] based on the related subscription.

- **IMS (IP Multimedia Subsystem)**: is the entity responsible for offering multimedia services' content and related signaling. IMS services are "based on an IETF defined session control capability which, along with multimedia bearers, utilizes the IP-Connectivity Access Network" [27].

In figure 1.3, the Red line represents normal data traffic between the mobile user and the destination (other mobile user, Web server, etc.). In case the two communicating parties are connected to the same operator, then the traffic will bounce at PGW without reaching PDN. The Blue line describes the path followed by user plane traffic of the services offered by the operator. The green line describes the path followed by the user generated control plane (signaling) traffic.

## 1.3.4. EPS AKA

A telecom operator offers restricted services to its customers and roaming users from other operators (if a roaming agreement exists with the user's home network). The operator has to filter out illegitimate users, and be able to bill those who benefit from the offered services. The procedure for identifying and authenticating users is shown in figure 1.4 [28].

The conceptual algorithm shown in figure 1.4 is used to define an abstract call flow having the following operations fulfilled sequentially:

- User identification
- Network authentication
- Mobile authentication

Network authentication (serving network authentication) was first added in EPS. The above call flow is valid for UMTS if network authentication is removed.

In UMTS and its predecessors, after authentication, the user can only assure "that he is connected to a serving network that is authorized by the user's HN to provide him services; this includes the guarantee that this authorization is recent" [29]. Security network authentication was considered unnecessary in UMTS, since there was an assumption of mutual trust among UMTS operators [30].

In EPS this assumption is considered not valid, thus serving network authentication was added to ensure that the serving network's identity is really confounded with what it is claiming. In other words, the user is connecting to the same network which its home operator has generated the AV (Authentication Vector) to.

The protocol responsible for identification and authentication in 3GPP mobile technologies is called AKA (Authentication and Key Agreement). More information on 3GPP EPS AKA is presented next.

EPS identifies subscribers permanently using a unique identifier called IMSI (International Mobile Subscriber Identity) [30]. Size of IMSI is 15-digits and divided into 3 digits for MCC (Mobile Country Code), 2 for MNC (Mobile Network Code) and 10 for MSIN (Mobile Subscriber Identification Number).

Capturing a user's permanent identifier transmitted over the air channel can be used to detect the user's current position in addition to user tracking and other privacy breaching attacks. 3GPP has tried to overcome these attacks by introducing a new temporary identifier, GUTI (Globally Unique Temporary UE Identity), to provide an unambiguous identification of the UE. GUTI is only relevant in the serving MME's area, thus it is sent by the network over a non-access stratum layer connection when confidentiality and integrity protected.

A user's GUTI is not expected to be constant over a long period of time, since it can then be used in user tracking instead of IMSI. GUTI might be updated in an attach accept message, tracking area update accept message or GUTI reallocation command. This modification is presented in using GUTI instead of previously used TMSI (Temporary Mobile Subscriber Identity).

As mentioned above, GUTI is only transmitted when the NAS layer connection is confidentiality and integrity protected. Confidentiality and integrity protection is enabled only after key sharing which is a late step in the access control mechanism. Conceptual steps in access control are: identification, authentication, authorization, key sharing, and finally enabling secure access. It can be seen from the discussed steps that identification occurs before establishing a secure connection, thus user identities are transmitted in plain text. Confidential identification is a very expensive task when compared to confidential data exchange in a security established connection, thus a compromise has to be made between additional costs resulting from ID hiding and high level of privacy. This dilemma faced the designers of GSM, UMTS, and finally LTE.

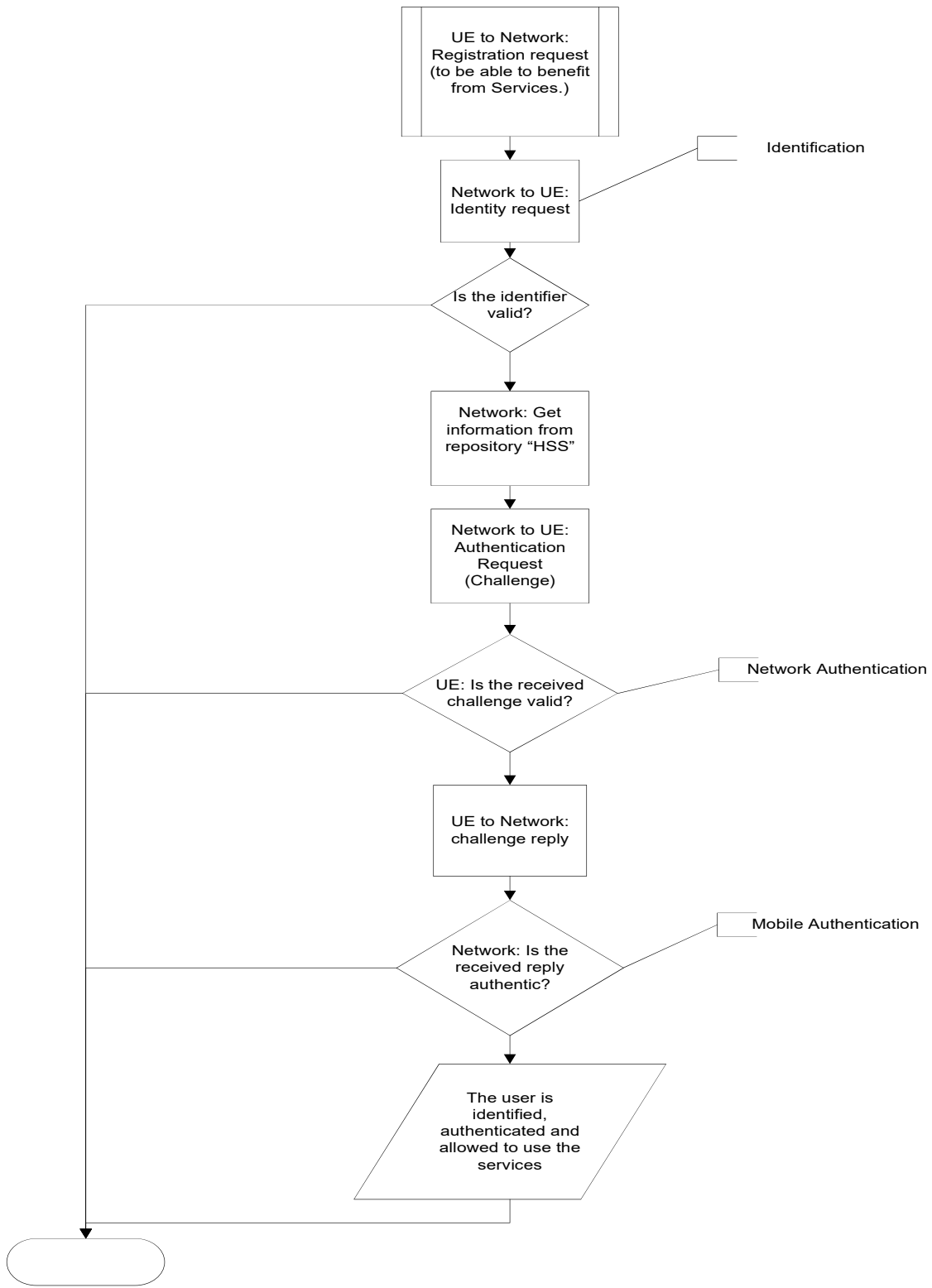

Figure 1.4 Identification and authentication conceptual algorithm

AKA (Authentication and Key Agreement procedure) is responsible for user identification, user authentication, network authentication and generation of master keys, which will be used to derive the keys used in deriving the confidentiality and integrity keys. This procedure includes UE (User Equipment), eNB (evolved NB), S-MME (Serving network's MME) and H-HSS (Home network HSS) as seen in figure 1.5.

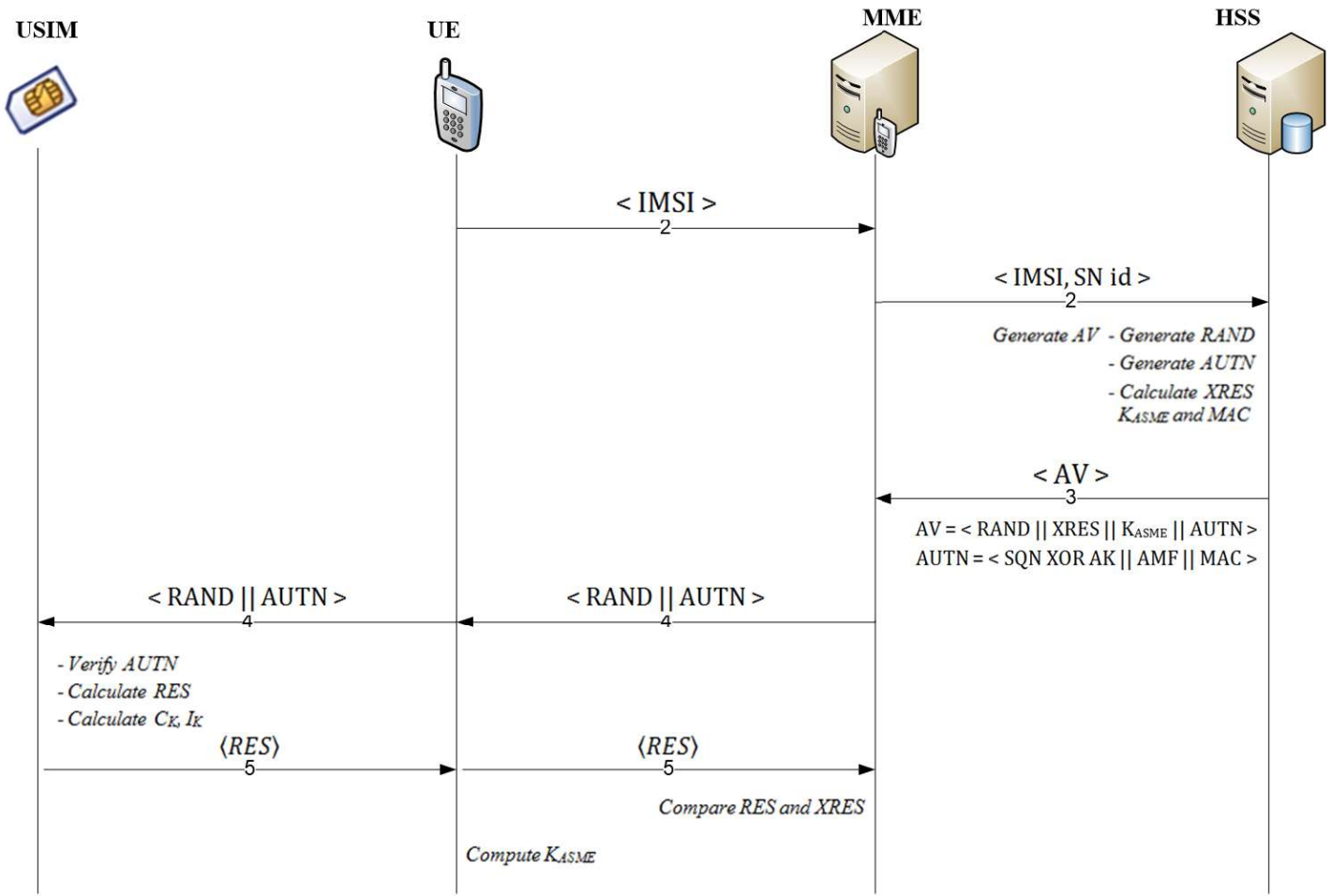

Figure 1.5 EPS authentication and key agreement protocol

EPS AKA procedure shown in figure 1.5 is as follows:

### 1.      UE → S-MME: NAS Attach Request (IMSI)

A user interested in connecting to a network, if the user has no previous temporary identifier, he has to identify himself by transmitting his permanent identifier (IMSI) in a NAS attach request. If the user has a temporary identifier from a previous connection, he can send his GUTI||LAI/RAI. The S-MME will contact the MME serving the LAI sent by the user, if it succeeded in retrieving GUTI/IMSI couplet it proceeds to step 2, else it requests the user to send his permanent identifier.

### 2.        S-MME → H-HSS:  Authentication Info Request (IMSI, SNID)

S-MME retrieves MCC||MNC from IMSI and route the request to H-HSS concatenated with the serving network's ID (IMSI||SNID). It then concatenates its ID to the request send by the user and forwards it to the corresponding HSS.

### 3.        H-HSS → S-MME:  Authentication Info Answer (RAND || XRES || KASME || AUTN)

The H-HSS fetches for IMSI/K/SQN triplet in its database. A random variable named RAND is generated. The remaining variables in the AV (Authentication Vector) are derived according to the following scheme:

- MAC = f1(K, AMF, SQN, RAND)
- AK = f5(K, RAND)
- AUTN = SQN xor (AK||AMF||MAC)
- CK = f3(K, RAND)
- IK = f4(K, RAND)
- KASME = KDF(CK, IK, SNID, (SQN xor AK))
- XRES = f2(K, RAND)
- AV = RAND||XRES|| KASME ||AUTN

AV is then sent back to S-MME.

### 4.        S-MME → UE: Authentication Request (RAND||AUTN)

MME forwards a challenge towards UE containing RAND and AUTN. UE verifies AUTN to authenticate legitimacy of the serving network. If the request is legitimate RES is generated in addition to the confidentiality and integrity keys.

**5.      *UE → S-MME: Authentication Reply (RES)***

UE replies to S-MME's challenge with RES. S-MME compares RES (from UE) and XRES (from H-HSS), if they are equal then the user is authenticated, keys will be derived, NAS security context will be established and S-MME sends eNB the needed keys to establish AS security context.

# 1.4.  Organization of this Report

The rest of this thesis is organized as follows. Chapter 2 proposes a resource optimizing mobile cloud architecture named OCMCA and rank this architecture against others found in literature to prove its superior performance. OCMCA's business model is also described in this chapter which is based on cloud federation. Chapter 3 proposes an authentication and single-sign-on protocol named EC-AKA3 which creates a trust context between the mobile user and her cloud provider passing through the mobile operator. Other privacy applications of OCMCA are presented in this chapter. Chapter 4 proves mathematically that cloud federation is financially feasible by itself in addition to its critical contribution to OCMCA's business model. This chapter also shows that "distance" is an important selection criterion to maintain the feasibility and profitability of cloud federation and proposes a federation manager to enforce this selection criterion. Finally chapter 5 summarizes our main conclusions, achievements and open issues for future research.

# CHAPTER 2  MOBILE CLOUD ARCHITECTURE: PERFORMANCE AND BUSINESS MODEL

## 2.1. Introduction

Mobile technology experienced radical changes in its concepts, aims and needs after introducing smartphones and 4G networks. Protocol traffic overhead appears to be no more a critical problem since mobile networks are achieving high throughputs and new technologies (such as 5G) are promising even higher rates. Mobile devices are not necessarily the slim clients that used to dominate the network ten years ago. Smartphones are powerful devices capable of processing majority of mobile applications, but have very limited power resources (e.g. battery) and it is not likely to change in the foreseen future. Mobile devices, either computation/storage-limited (slim client) or power-limited (Smartphone), need a technology that helps in decreasing power consumption, delay and user's cost without compromising the privacy, availability and mobility offered by mobile networks.

Mobile cloud computing is on-demand, dynamic and self-provisioned outsourcing of IT resources from a centralized service provider over a mobile access network. It is a technology that offloads resource-intensive applications from resource-limited mobile devices to be processed "somewhere else". This technology aims to decrease mobile's power consumption, allow complex applications to be managed from mobile devices and most importantly keep the expenses within the user's cost budget.

Many mobile applications are multicast-based in nature (such as: Google cloud messaging platform [31], multimedia applications and same-content delivery applications like: Google play, Apple store, etc.) but its physical implementation and mobile network's characteristics transform its communication into a group of unicast messages leading to unnecessary congestion, additional delay and more expensive fees. None of the mobile cloud architectures found in literature exploited this weakness but rather adopted it to become an inherited mobile cloud limitation.

In order to solve this problem and make mobile cloud computing more agreeable and efficient for both customers and operators, network core adaptations should be considered. In this chapter we discuss mobile cloud from telecommunication perspective by proposing an innovative architecture and interesting business model. This new architecture keeps the mobile operator at the center of mobile cloud computing and offers revenue-making business model that motivates operators to invest in this technology.

The rest of this chapter is organized as follows: In section 2.2 we survey existing mobile cloud architectures and applications found in literature. Section 2.3 presents the architecture requirements that are used to measure architecture performance and evaluate its suitability for various mobile cloud applications. It also presents the simulation environment, configuration and results which are analyzed to rank the architectures. Section 2.4 presents our proposed architecture which benefits from the mobile network to offer superior services as shown in its performance comparison with all other architectures. In this section, we also present our proposed architecture's business model which draws the guidelines to offer high profitability and penetration rates. Section 2.5 summarizes this chapter.

## 2.2. State-of-the-Art

Mobile cloud computing is relatively a new technology having lots of potential applications but no standardized architecture until nowadays. This motivated researchers to innovate different architectures each trying to emphasize on a certain requirement (such as mobility, power consumption, etc.) and select the optimal compromise for the others. In this section, we are going to survey existing mobile cloud architectures and applications found in literature.

### 2.2.1. Mobile Cloud Architectures

As a result of many innovative attempts, various mobile cloud architectures have been proposed each exploiting a technology (such as: ad hoc Wi-Fi, p2p, mobile networks, etc.) that helps in emphasizing on one of MCC's requirements. The mobile cloud computing architectures found in literature are:

### 2.2.1.1.          Cloud computing with mobile terminals

It is identical to the normal cloud computing architecture where computation is implemented in a remote cloud server (within the CSP's network), but the terminals in this case are mobile devices such as PDAs, Smart Phones, etc. Mobile devices connect to the Internet, in most of the cases, over an expensive connection such as LTE, UMTS and GPRS [32]. Connecting through Wi-Fi interface is also possible but it creates many concerns regarding mobility such as: network availability, handover, etc. An example on the above concerns is shown in the following use case: The user is accessing her CSP using a mobile device over a Wi-Fi network offered at the coffee shop she is currently in. Leaving the coffee shop makes her drop the connection and loses the cloud service. This architecture is shown in figure 2.1.

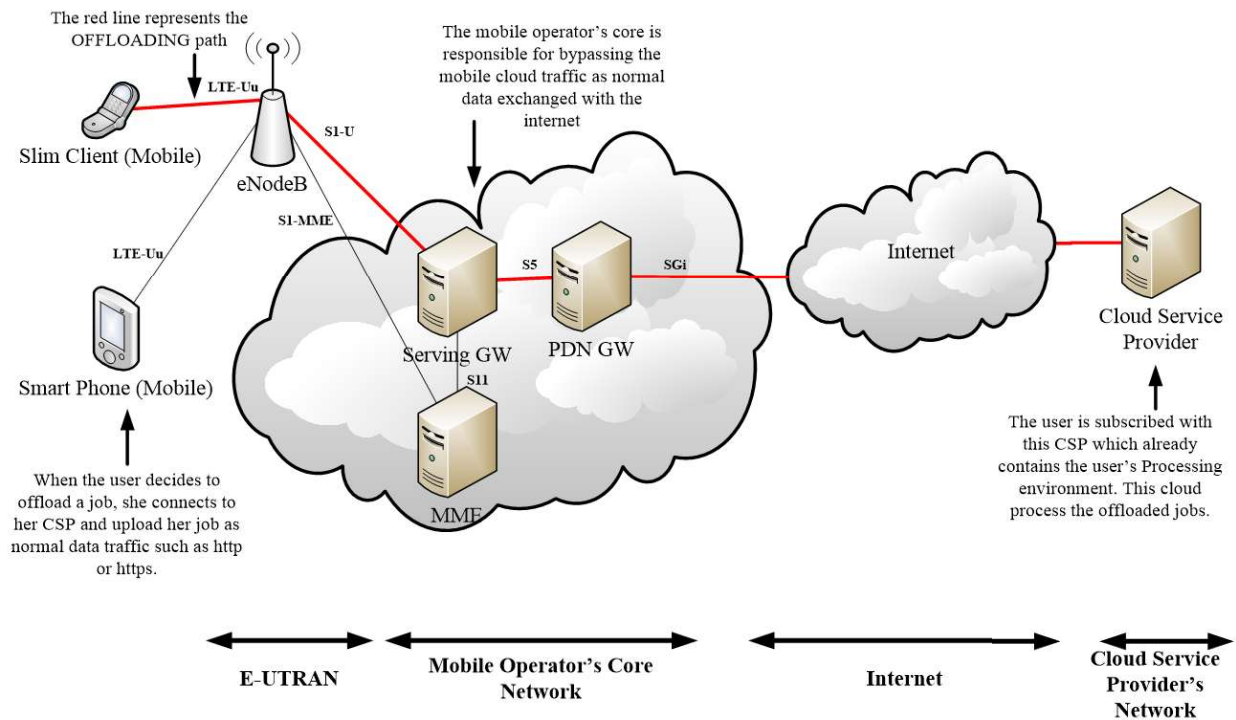

Figure 2.1"Cloud computing with mobile terminals" architecture

### 2.2.1.2.       **Virtual cloud computing provider**

This architecture proposes the creation of a virtual cloud from peer-to-peer connected mobile devices over an ad-hoc Wi-Fi connection to share processing burden [17]. P2P nodes participate only if found in the vicinity (Wi-Fi range) and interested in the processed data. After selecting the mobile devices participating in this virtual cloud, the job requestor divides the job into tasks and offloads each to a participant. After processing the task, the participant replies back with his processed data which is consolidated by the requestor to get the final result. These results are sent back to the participants. This architecture is shown in figure 2.2.

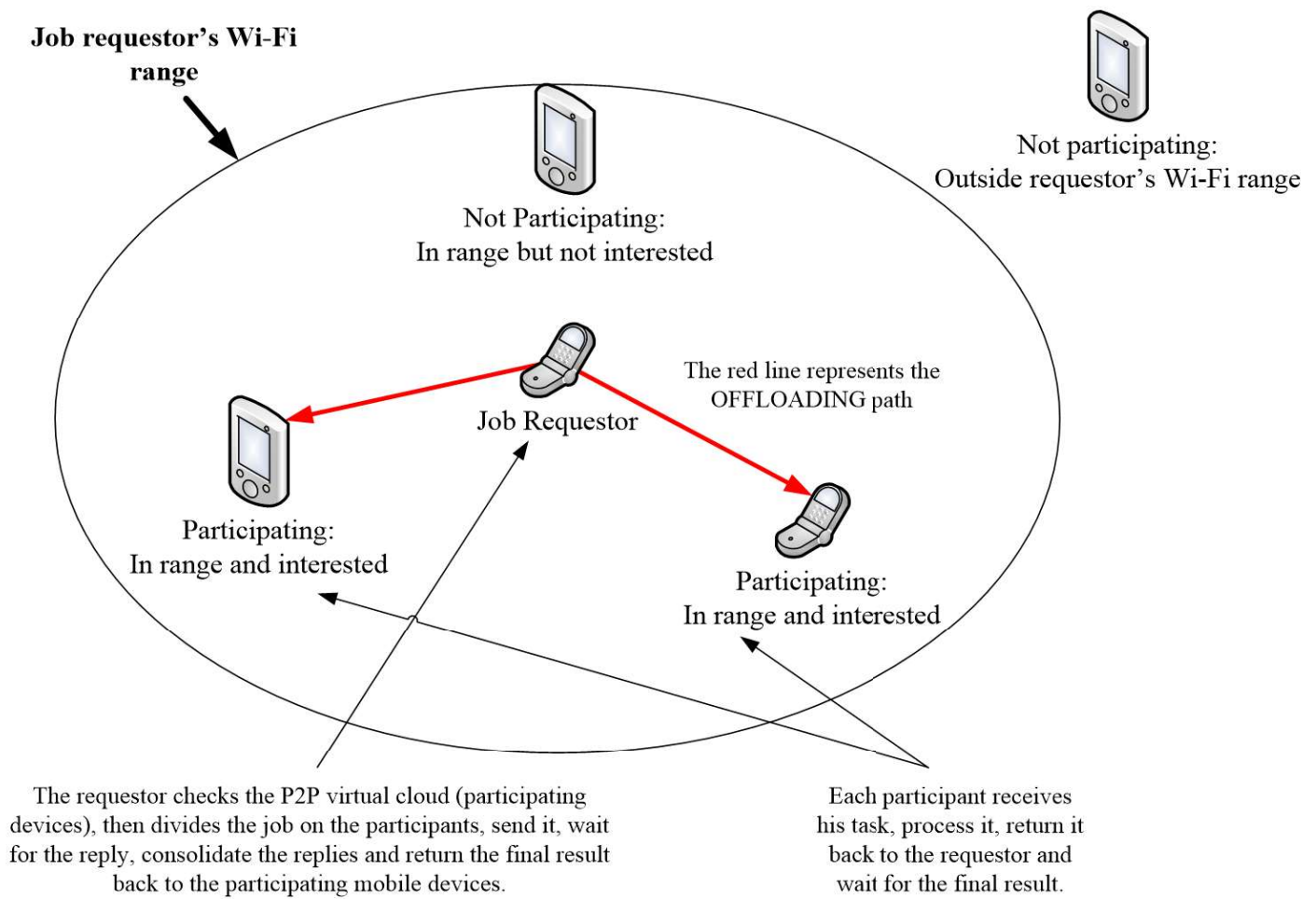

Figure 2.2"Virtual cloud computing provider" architecture

### 2.2.1.3.        Cloudlet

This architecture proposes installing cloudlet servers in high density areas (such as: coffee shops, malls, etc.) collocated with Wi-Fi hotspots [33]. The user connects to the cloudlet server and offloads its job to be processed. It contacts the user's cloud service provider to retrieve the user's environment through federation. Starting this point, future jobs will be processed at the cloudlet until getting disconnected (user leaves the hotspot coverage). This architecture is shown in figure 2.3.

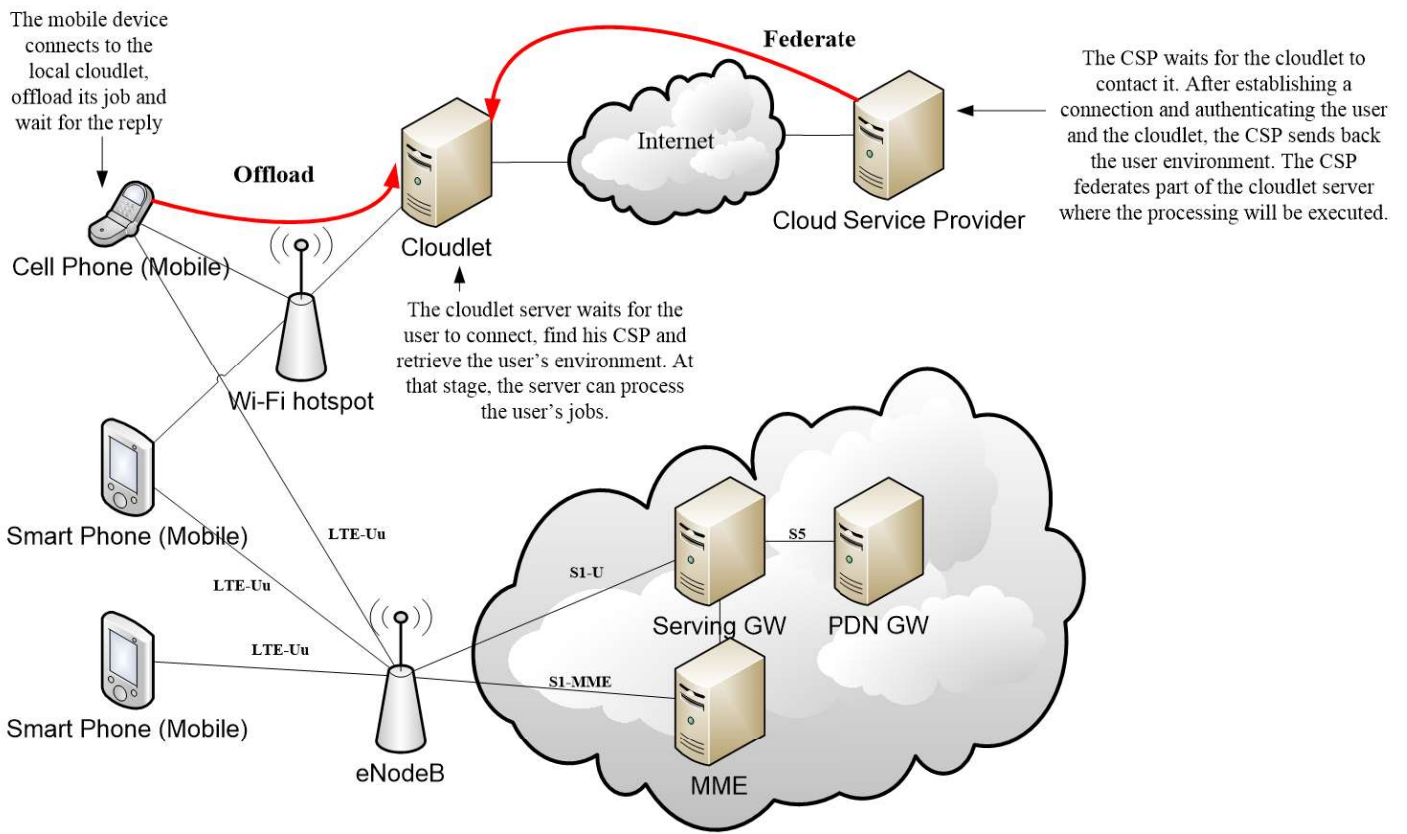

Figure 2.3"Cloudlet" architecture

### 2.2.1.4.        CloneCloud

It is having a clone of the mobile device running in cloud. This solution uses an application-level virtual machine that can partition an application and run one part on the mobile device and the other at the clone. This solution works at the application layer to decide which portion of an application should be offloaded to the cloud [32][34].

CloneCloud is an application layer solution, while all other architectures discussed above are physical layer and try to answer different questions (such as: how to connect to the cloud network? is user mobility ensured? how to ensure user confidentiality?, etc.). CloneCloud is complementary to the other architectures, thus will not be included in the performance comparison. Cuckoo [35], various Cloudlet versions [36][37], Spectra [38], Chroma [39], Hyrax [40], MMPI (Mobile Message Passing Interface) framework [41], MobiCloud[42] and MAUI [43] are offloading mechanisms and frameworks for dynamic selection of code partitions (pieces of the program) to be executed remotely similar to CloneCloud so will not be included in the performance comparison.

## 2.2.2. Mobile cloud Applications

Computation-intensive mobile applications are the best candidates for being transformed into mobile cloud applications. It has been shown in [44] that during 2012, games and messaging applications are the most popular from developer side (based on number of developed applications) but Facebook (76% of US Smartphone users), Google Maps (65.9%), Google Play (54.3%), Google Search (53.5%), Gmail (47.6%), YouTube (46.4%), Pandora Radio (42%), Apple iTunes (41%), Cooliris (38%) and Yahoo! Messenger (32%) are the most popular applications from user side (based on number of downloaded applications). The applications that that can benefit from multicast communication (Google Play, YouTube, Pandora Radio, Apple iTunes and Cooliris) constitute a majority of the top ranked mobile applications. In this section, we are going to present different computation-intensive mobile applications presented in literature [4] which is profiled and simulated in later sections. The applications are shown in table 2.1.

Table 2.1 Computation-intesive mobile applications

| Application group | Application |
|---|---|
| **OCR (Optical Character Recognition)**:Is processing an image to extract the found characters/text. Various applications can be implemented on the text after extraction, such as translation. | **Application 1**[4]: A foreign traveler tries to understand a street sign by capturing the image using his mobile, extracting the text using an OCR application and finally translating it to an understandable language. Another scenario for the same application was discussed in [17], where a foreign tourist is visiting a museum in South Korea. He is not able to understand an interesting exhibit written in Korean, so he captures the exhibit's image and tries to translate it using an OCR application. Since his mobile device is not able to process the captured image (due to limited or expensive resources such as RAM, Swap space and Internet connection), the application tries to scan nearby devices. Interested nearby devices create an ad hoc network and a virtual mobile cloud to process the image cooperatively. The extracted text is then translated to English. |
| **Natural language processing**: Is a useful tool for travelers to be able to communicate with locals. | **Application 2**[45]: Text-to-speech is an application allowing a mobile user having a file to be read to locals |
| **Crowd computing**: Is a method allowing different video recordings captured by different mobile devices to construct a single video covering an entire event [40]. | **Application 3 (Lost child)**[46]: A five years old John, is attending a parade with his parent in Manhattan. John goes missing and his parents report the incident to the Police, who send out an alert via a message to all mobile phones within two miles radius, requesting them to upload the parade images they have, to a server that only the police have access to. John is spotted in some images and his position was located, and he was reunited with his parents. This application requires high privacy as will be shown in section 3.3.2. Privacy is discussed in section 3.3.1. The participation invitation in this application is multicast/broadcast by nature. |
| | **Application 4 (Disaster relief)**[46]: Electronic maps become useless after a disaster, thus hindering disaster relief teams from performing rescue operations efficiently. Local citizens are asked to use their mobile phones to photograph disaster sites, and upload it to a central server [4]. The collected images are used to create a panoramic view of the sites, thus facilitating the navigation of the relief teams. This application requires high privacy similar to application 3. The participation invitation in this application is multicast/broadcast by nature. |
| **Sharing GPS/Internet data**[4]:Instead of reading common data from internet or GPS by multiple users, one user can download the data and share it with interested nearby devices through local-area or peer-to-peer networks. This application group helps in decreasing the cost and delay [40] resulting | **Application 5**[4]: Scans using Bluetooth for a co-located device which has a recent GPS reading. Instead of using the expensive GPS connection the needed information can be retrieved from co-located devices. The performance of this application augments in dense events such as: party, football match etc. Other format of this application is: "Traffic Lights Detector for Blind Navigation" [48]. The requestor's communication with other participants is multicast/broadcast by nature. |
| from downloading online information. | **Application 6**[17]: Instead of downloading a P2P file from the internet over an expensive interface (GPRS, UMTS, LTE etc), a mobile user scans using Bluetooth for a nearby device which has downloaded the needed file and retrieve it over a less expensive interface (Bluetooth). The requestor's communication with other participants is multicast/broadcast by nature. |
| | **Application 7**: We propose a new application which fits under this group. Bicycle fans gather yearly to watch "tour de France", a 23-day racing event [49], taking place mainly in France and other nearby countries. Fans and racers are interested in knowing the instantaneous detailed ranking of racers with additional information (the time difference between the first racer and other contestants, velocity etc.). Each user needs to receive the data on his device, which is transmitted to all application users. The transmitted data is similar to all the users and uses multicast/broadcast by nature. |
| **Crowdsensing**: Are used to share timestamped sensor readings, such as GPS, accelerometer, light sensor, microphone, thermometer, clock, and compass. | **Application 8**[4]: Queries the users located in 1 mile radius to get the average temperature of nodes within a mile. The requestor's communication with other participants is multicast/broadcast by nature. |
| | **Application 9**[40]: Traffic reporting can be implemented by querying the velocity distribution of all nodes within half a mile of the next highway on the current route. The requestor's communication with other participants is multicast/broadcast by nature. |
| **Multimedia search**[4]: Mobile devices store many types of multimedia content such as videos, photos, and music which can be shared by other users. | **Application 10**[40]: Multimedia files can be searched in the contents of nearby mobile devices. The query in this application is multicast/broadcast by nature. |
| **Social networking**[4]:Sharing user content with friends on social media facilitates automatic sharing and P2P multimedia access, thus reducing the need for huge servers to manage this amount of data [40]. | **Application 11**[40]: Integrating Facebook with mobile cloud to share the active files using mobile interfaces. |
| **Crowdsourcing**: Is outsourcing tasks, used to be executed by machines or employees, into an external set of people [50][51]. | **Application 12**: "Social search engine" is one of the most popular crowdsourcing application theme which focuses on answering context-related questions using human-help (crowd) instead of or in complementary with search engines [52][53]. Chacha [54] and Aardvark [55] are two popular crowdsourcing applications that have similar objectives, but for detailed discussion, application 12 will represent Chacha. This application requires high privacy as shown in section 3.3.1. |
| | **Application 13**: "Crowdsourced location-based service" is a method to fetch the recommendations about certain location based categories posted by people with taste and interest similar |

| | to the requester. Foursquared[56] is a popular application offering crowdsourced location based service. This application requires high privacy as shown in section 3.3.1. |
|---|---|
| **Mobile augmented reality [57][58]:**Is the creation of an information-intensive virtual world confounded with the physical environment allowing the user to "display related information, to pose and resolve queries and to collaborate with other people" [57]. Mobile augmented reality (MAR) is running AR applications in the mobile environment | **Application 14 [59]:** "Google glasses" is a wearable computing platform which allows the person wearing those glasses to view the real word in addition to supplementary requested information. It can be used also to execute certain jobs such as sending messages, translating etc. |
| **Wearable computing [57][58]:**Accessing information generated (sensor) or received (downloaded) using mini-electronic devices having very limited battery, storage and computation resources. | **Application 15:** A tool to measure the blood pressure, heart rate and other vital signs of the user and upload these information to be presented in real-time to the doctor. |

There are other mobile cloud applications found in literature but are still under investigation, such as Cloud gaming [60][61][62][63], Mobile learning [60][64][65][66][67], and Mobile healthcare [60][68][69]. These applications cannot be profiled, due to the variety of proposed solutions and optimization techniques, thus will not be included in the simulation results.

# 2.3. Architectures evaluation

In this section, we are going to prove that none of the architectures found in the literature is suitable for all applications and none can be selected as optimal. To do so, we start by showing the application requirements which is used to evaluate the performance of each architecture. We then profile each application into numerical values (application configuration) which is used in the simulation. The network is then configured into numerical values and finally the performance of all the architectures is shown and analyzed.

## 2.3.1. Architecture Requirements

MCC (Mobile Cloud Computing) was developed to respond to mobile devices' needs for a technology that helps in decreasing power consumption, delay and user's cost without compromising the privacy, availability and mobility offered by mobile networks, as shown in section 2.1. When designing a mobile cloud architecture several, requirements should be taken into consideration which represents the aim behind using MCC. Similarly, these requirements are used as metrics to evaluate the performance and compare the studied architectures. The requirements are divided into quantifiable and non-quantifiable as shown next:

- **Quantifiable requirements**: can be calculated in numerical metrics. The quantifiable requirements are:

  - • **Cost**: Financial cost due to network usage. Cost is calculated in terms of "financial units" relative to the mobile fees. An architecture better suits the application if it achieves lower cost.

  - • **Delay:** Transmission, propagation and processing delay. In case of in-house processing, it is the time between starting the execution and finishing the job. In case of offloading, it is the time between sending the first offloaded bit till receiving the last reply bit. Delay is calculated in milliseconds (ms). An architecture better suits the application if it achieves lower delay.

  - • **Power consumption**: Power consumed by a mobile device during processing, transmission and waiting. It is calculated in power units. An architecture better suits the application if it achieves lower power consumption.

- **Non-quantifiable requirements**: cannot be calculated in numerical metrics, but will be represented by subjective values. The non-quantifiable requirements are:

  - • **Privacy:** Privacy of data during transmission and processing. Privacy is considered high, if no private personal data is stored or processed in non-trusted devices. It is considered medium if private personal data (such as the temperature read by a mobile's sensor) is stored or processed in non-trusted devices, and considered low if private personal data is stored or processed in non trusted devices and can be correlated to the user's id. A cloudlet server is considered non-trusted since it is not under the supervision of a trusted provider. An architecture better suits the application if it achieves higher privacy. Architectures get disqualified if it is not able to satisfy the privacy requirements of an application.

- • **Mobility:** The expected distance to be covered by a user before disconnecting from a service. Cloudlet servers can be accessed through local APs (Access Points) only, then this very short range hinders the device's online mobility especially that we don't expect LAN coverage to be continuous. We consider that Cloudlet offers very low mobility. In virtual cloud, the user has more mobility freedom since not bounded to the fixed position of the AP, but bounded to the group forming the virtual cloud he is connected. We consider this architecture offers medium mobility. The user in "Cloud computing with mobile terminals" has the freedom to move in the entire region covered by his or any another operator (using roaming). We consider this architecture offers high mobility. Since mobility is one of MCC's requirements, we consider that a studied architecture better suits an application if it achieves higher mobility.

- • **Scalability:** Mobile applications downloaded in 2013 range between 56 and 82 billion and it is expected to reach 200 billion in 2017 [44]. Mobile cloud architectures are expected to experience high penetration rates which leave the researchers in front of a critical scalability issue. The architecture should be able to serve millions of users, handle the incremental traffic mobile networks are currently facing and most importantly offer a satisfying coverage in cities and major villages. Each architecture's scalability is studied separately:
  - o **Cloud computing with mobile terminals:** This architecture offers excellent coverage and routes all the MCC jobs to be executed at CSP's premises. CSPs are able to process huge amount of jobs and extend his capability by federating some resources from other cloud providers. The only bottleneck is at transmission which might be overwhelmed with the incremental traffic if no newer mobile technology is implemented. This architecture requires no investment in additional physical devices and able to support current traffic rates. This architecture is considered to have high scalability.

o **Virtual cloud computing provider:** This architecture requires no investment in additional physical devices, but will have difficulties (delay and power drainage) in processing very complex applications since the participating mobile devices are sharing the load. A user interested in processing more complex applications should team up with more users, users with powerful mobile devices and/or upgrade his devices. This architecture is considered to have high scalability.

o **Cloudlet:** This architecture requires the investment of random business owners in deploying physical devices (cloudlet servers) even with the absence of a valid business model. Cloudlets will face low utilization rates, over-investment and high rejection rates due to the traffic commuting. As residential areas are usually apart from business areas, the cloudlet servers in one area will be severely under-utilized throughout the period the users have commuted to other locations. This requires over-investment to offer acceptable services in different areas. If some areas are not well equipped, customer rejection rates will be elevated. This architecture is considered to have low scalability.

- • **Multicast-capable**: Boolean value specifying whether the studied architecture is capable of handling multicast traffic efficiently (not transforming it into bulk of unicast traffic). Since the majority of the top ranked mobile applications are multicast-based, having a multicast-capable mobile cloud architecture is of critical importance.

## 2.3.2. Application configuration

In this section, we profile the mobile cloud applications into numerical values which are used, in the simulation shown later, as application configuration. Each presented application is profiled using the following parameters:

- **Uploaded data**: Size of the uploaded data which will be processed. Unit is byte.
- **Processing**: The processing needed in terms of processing units to generate the response. Unit is "processing unit".
- **Downloaded data**: Size of the processed data downloaded. Unit is byte.
- **Resource sharing**: Boolean value represents whether the client requires accessing the resources (GPS, Internet, etc.) of the cloud server (cloudlet or other mobile devices generating a virtual cloud).

To better simulate the architectures, each application is replaced by its profile and a sample job will be executed. Application profiling is shown in table 2.2.

Table 2.2 Application profiling

| Application | Uploaded Data (KB) | Downloaded Data (KB) | Processing (processing units) | Resource sharing |
|---|---|---|---|---|
| 1 | 1,000 | 1 | 100,000 | False |
| 2 | 1 | 1,000 | 1,000 | False |
| 3 & 4 | 1 | 100,000 | 10,000 | False |
| 5 | 1 | 1 | 100 | True |
| 6 & 10 | 1 | 10,000 | 1,000 | True |
| 7 | 1 | 1,000 | 1,000 | True |
| 8 & 9 | 1 | 1 | 100,000 | True |
| 11 | 1,000 | 1,000 | 1,000 | True |
| 12 & 13 | 1 | 1 | 1,000 | False |
| 14 | 1,000 | 1 | 100,000 | True |
| 15 | 10 | 1 | 10,000 | False |

### 2.3.3. Network configuration

In this section, we show the network configuration we used to simulate various mobile cloud architectures and applications. The network is composed of two parts (end devices and transport network) and the configuration of each part is shown separately.

Transport network configuration is based on LTE experimental values reported in [70][71] which are shown in Table 2.3.

Table 2.3 Transport network configuration

| Parameter | Value |
|---|---|
| Air Interface Latency | 8 ms |
| eNB Processing Latency | 3 ms |
| Air Interface Upload bandwidth | 0.35 Mbps |
| Air Interface Download bandwidth | 36 Mbps |
| "Backhaul + Core" Delay | 5 ms |
| "Backhaul + Core" Upload bandwidth | 100 Mbps |
| "Backhaul + Core" Download bandwidth | 100 Mbps |
| S-GW/P-GW Processing Latency | 2 ms |
| Internet Latency | 25 ms |
| Internet Upload bandwidth | 100 Mbps |
| Internet Download bandwidth | 100 Mbps |

End device configuration is retrieved from [72][73] and shown in Table 2.4.

Table 2.4 End device configuration

| Parameter | Value |
|---|---|
| Mobile device (UE) processing speed | 1000 unit per second |
| Cloud processing speed (100X faster than UE) "Assumed" (depends on the used UE and VM) | 100,000 unit per second |
| Cloudlet Server processing speed [73] | 5,000 unit per second |
| UE's power cost for sending 1 bit over Wi-Fi | 1 power units |
| UE's power cost for sending 1 bit over LTE [72] | 23 power units |
| UE's power cost while waiting for 1 second [73] | 30,000,000 power units |
| UE's power cost while computing over 1 second [73] | 80,000,000 power units |
| UE's cost for sending 1 bit over Wi-Fi to cloudlet | 1 financial unit |
| UE's cost for sending 1 bit over LTE (recorded in Lebanon on | 1 - 100 financial units |
| 2014, these scores are country/operator related). Some mobile operators offer "lump sum" fees for their local users and elevated expenses for roaming users which are represented by the following range. | (1 for local user 100 for roaming user) |
| Wi-Fi bandwidth (upload and download) | 100 Mbps |

The abstract "power cost" values are used to make the simulation "device-independent". To give an insight about real values, HP iPAQ PDA with a 400-MHz Intel XScale processor has the following scores:

- UE's power cost while waiting for 1 second: 0.3 W [73].
- UE's power cost while computing for 1 second: 0.9 W [73].

Although these values might be different in other mobile devices, but the abstract values maintain the proportion. Note: We have considered the mobile device's power consumption during "LTE_ACTIVE short drx" state and discarded other states (such as: LTE_idle, LTE_active long drx, etc.).

## 2.3.4. Simulation Results

In this section, we show the simulation results of the mobile cloud architectures and applications presented earlier. The simulation used specially crafted C# programs based on the configuration parameters also shown earlier. The performance of each architecture-application couplet is evaluated based on the architecture requirements. The programs use internal clocking that makes them independent from the platform's clock and scheduler's performance thus creates the same result on any used platform (Hardware, Operating System, etc.). We have assumed in our calculation that cloudlet servers and interested mobile users are always found, so architectures are not disqualified based on applications' mobility requirements.

Not all applications discussed in section 2.2.2 can be executed in-house (by the mobile device itself without using any connection such as: Wi-Fi, Mobile data, Bluetooth and GPRS), thus the performance of standalone mobile devices are shown in table 2.5.

Table 2.5 Performance evaluation of standalone mobile devices

| | Quantifiable Requirements (Lower values are better) | | | Non-quantifiable Requirements (Higher values are better) | | | |
|---|---|---|---|---|---|---|---|
| Appl. | Cost (financial units) | Delay (ms) | Power consumption ($10^6$ power units) | Privacy | Mobility | Scalability | Multicast Capable? |
| 1 | 0 | 100,000 | 8000 | High | High | - | - |
| 2 | 0 | 1,000 | 80 | High | High | - | - |

Processing jobs in-house definitely ensures optimal scores when it comes to cost, privacy and mobility; but drastically fallback in delay and power consumption. For this reason mobile cloud computing is needed. The simulation results for "Cloud computing with mobile terminals" mobile cloud architecture are shown in table 2.6.

Table 2.6 Performance evaluation of the mobile cloud architecture "Cloud computing with mobile terminals"

| | Quantifiable Requirements (Lower values are better) | | | Non-quantifiable Requirements (Higher values are better) | | | |
|---|---|---|---|---|---|---|---|
| Appl. | Cost ($10^6$ financial units) | Delay (ms) | Power consumption ($10^6$ power units) | Privacy | Mobility | Scalability | Multicast Capable? |
| 1 | > 1.001 < 100.1 | 3963 | 142.033 | High | High | High | No |
| 2 | > 1.001 < 100.1 | 145 | 27.433 | High | High | High | No |
| 3 & 4 | > 100.001 < 10000.1 | 245 | 2440.543 | High | High | High | No |
| 5 | > 0.002 < 0. 2 | 89 | 2.896 | High | High | High | No |
| 6 & 10 | > 10.001 < 1000.1 | 165 | 246.553 | High | High | High | No |
| 7 | > 1.001 < 100.1 | 145 | 27.433 | High | High | High | No |
| 8 & 9 | > 0.002 < 0.2 | 1088 | 32.86 | High | High | High | No |
| 11 | > 2 < 200 | 3020 | 112.333 | High | High | High | No |
| 12 & 13 | > 0.002 < 0.2 | 98 | 3.166 | High | High | High | No |

| 14 | > 1.001 < 100.1 | 3963 | 142.033 | High | High | High | No |
| 15 | > 0.011 < 1.1 | 214 | 6.853 | High | High | High | No |

The cost range shown in table 2.6 represents the range between local and roaming users. This architecture cannot overcome in-house processing in cost, but has achieved considerable enhancements in delay and power consumption for both applications. The simulation results for "Virtual cloud computing provider" mobile cloud architecture are shown in table 2.7.

Table 2.7 Performance evaluation of the mobile cloud architecture "Virtual cloud computing provider"

| Appl. | Quantifiable Requirements (Lower values are better) | | | Non-quantifiable Requirements (Higher values are better) | | | |
|---|---|---|---|---|---|---|---|
| | Cost ($10^6$ financial units) | Delay (ms) | Power consumption ($10^6$ power units) | Privacy | Mobility | Scalability | Multicast Capable? |
| 1 | 0 | 10010 | 801.001 | Medium | Medium | High | No |
| 2 | 0 | 110 | 9.001 | Medium | Medium | High | No |
| 3 & 4 | Disqualified | | | | | | |
| 5 | 0.02 | 10 | 0.802 | Medium | Medium | High | No |
| 6 & 10 | 100.01 | 114 | 18.001 | Medium | Medium | High | No |
| 7 | 10.01 | 110 | 9.001 | Medium | Medium | High | No |
| 8 & 9 | 0.02 | 10000 | 800.002 | Medium | Medium | High | No |
| 11 | Disqualified | | | | | | |
| 12 & 13 | Disqualified | | | | | | |
| 14 | 10.01 | 10010 | 801.001 | Medium | Medium | High | No |
| 15 | 0 | 1000 | 80.011 | Medium | Medium | High | No |

It is shown in figure 2.7 that "Virtual cloud computing provider" has been disqualified in different applications (App. 3, 4, 11, 12 and 13) since failed to meet applications' privacy constraints (failed to offer the privacy level required by an application). For the remaining applications, this architecture overcomes "Cloud computing with mobile terminals" in delay and power consumption (for App. 2, 5, 6, 7 and 10) and fails to compete in the same requirements for applications 1, 8, 9, 14 and 15. We can conclude that this architecture suits applications 2, 5, 6, 7 and 10 more than "Cloud computing with mobile terminals". These results will be analyzed thoroughly in section 2.3.5. Table 2.8 studies the third architecture "Cloudlet" against various applications.

Table 2.8 Performance evaluation of the mobile cloud architecture "cloudlet"

| Appl. | Quantifiable Requirements (Lower values are better) | | | Non-quantifiable Requirements (Higher values are better) | | | |
|---|---|---|---|---|---|---|---|
| | Cost ($10^6$ financial units) | Delay (ms) | Power consumption ($10^6$ power units) | Privacy | Mobility | Scalability | Multicast Capable? |
| 1 | 1.001 | 20010 | 601.001 | Low | Low | Low | No |
| 2 | 1.001 | 210 | 7.001 | Low | Low | Low | No |
| 3 & 4 | Disqualified | | | | | | |
| 5 | 0.002 | 20 | 0.602 | Low | Low | Low | No |
| 6 & 10 | 10.001 | 214 | 16.001 | Low | Low | Low | No |
| 7 | 1.001 | 210 | 7.001 | Low | Low | Low | No |
| 8 & 9 | 0.002 | 2000 | 600.002 | Low | Low | Low | No |
| 11 | Disqualified | | | | | | |
| 12 & 13 | Disqualified | | | | | | |
| 14 | 1.001 | 20010 | 601.001 | Low | Low | Low | No |
| 15 | 0.11 | 2000 | 60.011 | Low | Low | Low | No |

It is shown in figure 2.8 that "Cloudlet" achieves low scores in privacy, mobility, scalability and delay but relatively good scores in power consumption.

## 2.3.5.  Result Analysis

In this section, we deduce the strengths and weaknesses of each architecture based on the simulation results. We then rank the best performing architectures across requirements and prove that no mobile cloud architecture is suitable for all applications. We start by comparing the architectures.

### 2.3.5.1.        Cloud computing with mobile terminals vs. Virtual cloud computing provider

When comparing the delay and power consumption results of these two architectures, we find that:

- "Virtual cloud computing provider" overcomes "Cloud computing with mobile terminals" in application: 2, 5, 6, 7 and 10.
- "Cloud computing with mobile terminals" overcomes "Virtual cloud computing provider" in application: 1, 8, 9, 14 and 15.
- "Virtual cloud computing provider" fails in satisfying the privacy requirements for application: 3, 4, 11, 12 and 13.

Applications 2, 5, 6, 7 and 10 have "low processing" ($\leq 1000$ processing units) as a common property. "Cloud computing with mobile terminals" does not overcome "Virtual cloud computing provider" in any application with this property. We deduce that "Virtual cloud computing provider" suits applications with "low processing".

Applications 1, 8, 9, 14 and 15 have "high processing" ($\geq 10,000$ processing units) as a common property. "Cloud computing with mobile terminals" fail to overcome "Virtual cloud computing provider" in any application with this property. We deduce that "Cloud computing with mobile terminals" suits applications with "high processing".

"Virtual cloud computing provider" fails to provide the privacy requirements for applications 3, 4, 11, 12 and 13. In addition to that, this architecture scores lower than "Cloud computing with mobile terminals" on mobility. Both architectures have good scalability but fail to take advantage of multicast traffic.

### 2.3.5.2.        Cloud computing with mobile terminals vs. Cloudlet

When comparing the delay results of these two architectures, we find that "Cloud computing with mobile terminals" overcomes "Cloudlet" for all applications except one (App. 5).

As for power consumption "Cloudlet" overcomes "Cloud computing with mobile terminals" in applications 2, 5, 6, 7 and 10 which are the applications with "low processing". We deduce that "cloudlet" suits applications with "low processing" and we confirm our previous deduction that "Cloud computing with mobile terminals" suits applications with "high processing".

### 2.3.5.3.       Virtual cloud computing provider vs. Cloudlet

When comparing the delay results of these two architectures, we find that "Virtual cloud computing provider" overcomes "Cloudlet" for all applications except applications 8 and 9. These two are the only applications having high processing (100,000 processing units) and very small uploaded data (1KB).

As for power consumption "Cloudlet" overcomes "Virtual cloud computing provider" in all applications. We deduce that "cloudlet" results in lower power consumption than "Virtual cloud computing provider".

### 2.3.5.4.       Architecture Ranking

In order to rank the cloud architectures for each application, many grading schemes can be used. We are going to use five different schemes just to show that for different preferences no mobile cloud architectures can be considered optimal. The grading schemes are:

- The first grading scheme orders architectures based on delay. The architectures are ranked in table 2.9, where smaller rank is better.
- The second grading scheme orders architectures based on power consumption. The architectures are ranked in table 2.10, where smaller rank is better.
- The third grading scheme orders architectures based on cost. The architectures are ranked in table 2.11, where smaller rank is better.
- The fourth grading scheme orders architectures based on privacy and mobility. The architectures are ranked in table 2.12, where smaller rank is better.
- The fifth grading scheme orders architectures based on scalability. The architectures are ranked in table 2.13, where smaller rank is better.

Table 2.9 Ranking cloud architectures using first scheme

| Appl. | Executing in-house | Cloud computing with mobile terminals | Virtual cloud computing provider | Cloudlet |
|---|---|---|---|---|
| 1 | 4 | 1 | 2 | 3 |
| 2 | 4 | 2 | 1 | 3 |
| 3 & 4 | - | 1 | X | X |
| 5 | - | 3 | 1 | 2 |
| 6 & 10 | - | 2 | 1 | 3 |
| 7 | - | 2 | 1 | 3 |
| 8 & 9 | - | 1 | 3 | 2 |
| 11 | - | 1 | X | X |
| 12 & 13 | - | 1 | X | X |
| 14 | - | 1 | 2 | 3 |
| 15 | - | 1 | 2 | 3 |

"Cloud computing with mobile terminals" becomes more competitive against other architectures in delay as the processing becomes more complex. Oppositely, "Virtual cloud computing provider" becomes more competitive against other architectures in its delay results as the processing becomes lighter. "Cloudlet" architecture offers higher delay for all applications, which is considered a drawback.

Table 2.10 Ranking cloud architectures using second scheme

| Appl. | Executing in-house | Cloud computing with mobile terminals | Virtual cloud computing provider | Cloudlet |
|---|---|---|---|---|
| 1 | 4 | 1 | 3 | 2 |
| 2 | 4 | 3 | 2 | 1 |
| 3 & 4 | - | 1 | X | X |
| 5 | - | 3 | 2 | 1 |
| 6 & 10 | - | 3 | 2 | 1 |
| 7 | - | 3 | 2 | 1 |
| 8 & 9 | - | 1 | 3 | 2 |
| 11 | - | 1 | X | X |
| 12 & 13 | - | 1 | X | X |
| 14 | - | 1 | 3 | 2 |
| 15 | - | 1 | 3 | 2 |

"Cloud computing with mobile terminals" becomes more competitive against other architectures in power consumption as processing becomes more complex. Oppositely, "Cloudlet" becomes more competitive against other architectures in power consumption as the processed data becomes lighter."Virtual cloud computing provider" architecture offers higher power consumption for all applications, which is considered a drawback.

Table 2.11 Ranking cloud architectures using third scheme

| Appl. | Executing in-house | Cloud computing with mobile terminals (local user) | Cloud computing with mobile terminals (Roaming user) | Virtual cloud computing provider | Cloudlet |
|---|---|---|---|---|---|
| 1 | 1 | 2 | 4 | 1 | 2 |
| 2 | 1 | 2 | 4 | 1 | 2 |
| 3 & 4 | - | 1 | 2 | X | X |
| 5 | - | 1 | 4 | 3 | 1 |
| 6 & 10 | - | 1 | 4 | 3 | 1 |
| 7 | - | 1 | 4 | 3 | 1 |
| 8 & 9 | - | 1 | 4 | 3 | 1 |
| 11 | - | 1 | 2 | X | X |
| 12 & 13 | - | 1 | 2 | X | X |
| 14 | - | 1 | 4 | 3 | 1 |
| 15 | - | 2 | 4 | 1 | 3 |

"Cloud computing with mobile terminals" achieves best cost performance for local users across all applications except (1, 2 and 15), but worst performance for roaming users across all applications. "Cloudlet" has good performance with applications 5,6,7,8,9,10 and 14 while "Virtual cloud computing provider" performs better with applications 1, 2 and 15.

Table 2.12 Ranking cloud architectures using fourth scheme

| Appl. | Executing in-house | Cloud computing with mobile terminals | Virtual cloud computing provider | Cloudlet |
|---|---|---|---|---|
| 1 | 1 | 2 | 3 | 4 |
| 2 | 1 | 2 | 3 | 4 |
| 3 & 4 | - | 1 | X | X |
| 5 | - | 1 | 2 | 3 |
| 6 & 10 | - | 1 | 2 | 3 |
| 7 | - | 1 | 2 | 3 |
| 8 & 9 | - | 1 | 2 | 3 |

| **11** | - | 1 | X | X |
|--------|---|---|---|---|
| **12 & 13** | - | 1 | X | X |
| **14** | - | 1 | 2 | 3 |
| **15** | - | 1 | 2 | 3 |

Similarly, it is most private to execute a job in-house, followed by a trusted service provider (Cloud computing with mobile terminals), then a collection of users share the same interest in the processed job (Virtual cloud computing provider) and finally the least privacy offering architecture is "Cloudlet" where data is processed in a server not under the monitoring of a trusted party.

Table 2.13 Ranking cloud architectures using fifth scheme

| **Appl.** | **Cloud computing with mobile terminals** | **Virtual cloud computing provider** | **Cloudlet** |
|-----------|-------------------------------------------|--------------------------------------|--------------|
| **1** | 1 | 1 | 3 |
| **2** | 1 | 1 | 3 |
| **3 & 4** | 1 | X | X |
| **5** | 1 | 1 | 3 |
| **6 & 10** | 1 | 1 | 3 |
| **7** | 1 | 1 | 3 |
| **8 & 9** | 1 | 1 | 3 |
| **11** | 1 | X | X |
| **12 & 13** | 1 | X | X |
| **14** | 1 | 1 | 3 |
| **15** | 1 | 1 | 3 |

"Cloud computing with mobile terminals" and "Virtual cloud computing provider" are considered to have higher scalability than "Cloudlet" which requires investment by third parties even without a valid business model. For "Cloudlet" to be a well deployed mobile cloud architecture, huge number of servers should be implemented to offer suitable coverage.

Based on the shown comparisons and grading schemes, we can deduce that no architecture best suits all applications across all requirements. These comparisons are studied from user's perspective i.e. to have lower cost, power consumption and delay in addition to higher mobility, scalability and privacy. Although the mobile operator can impose the greatest influence in implementing MCC, architecture comparison and selection, in literature and throughout this chapter, were focusing on users' needs while neglecting the operator's concerns. "Cloudlet" and "Virtual cloud computing provider", for example, eliminate the use of the mobile operator. "Cloud computing with mobile terminals" significantly increase the traffic leading to higher rejection rates and lower quality for real-time applications, such as voice calls, which are the operator's major profit.

The problem in all existing mobile cloud architectures (previously surveyed) is that they discard the financial incentive behind MCC and fail to take advantage of the major traffic type (multicast) generated by computation-intensive mobile applications. We propose, in the coming section, a mobile cloud architecture satisfying both concerns (users' and operators') enforced with a valid business model that encourages the investment of stakeholders (such as: mobile vendors, mobile operators and cloud providers) in this architecture.

## 2.4. The proposed architecture

The decision makers in the technology industry are biased by financial and economic motivations which make their decisions, in many cases, technology unfair. An adopted solution should be technologically and financially feasible and introduces a revenue-making business model. Even a good solution doesn't make it to implementation phase if it discards these motivations, especially if a better solution exists and focuses on promoting economic incentives?

The slow progress in deploying IPv6 is a very well-known example on economic effects and market statics that goes even beyond major technology players who are pushing toward this migration. Another example is Fujitsu which has developed a data transfer protocol that overcomes "UDP" and "TCP's" speed by 30 times [74]. Although Fujitsu has achieved impressive breakthrough, they are not expecting it to replace "TCP" or "UDP" because of the absence of a valid business model that persuade the huge investment needed to perform this change (new TCP/IP stacks, APIs, application upgrades, etc.).

We have shown in a previous section that the majority of the top ranked mobile applications are multicast-based, and none of the mobile cloud architectures found in literature was designed to take advantage of this mobile traffic characteristic. The weakness in existing architectures is that they tried to meet MCC requirements through network edge solutions (focusing on network edge entities without any modification to the network core) by either trying to eliminate the use of mobile networks (such as "Cloudlet" and "Virtual cloud computing provider") or abstracting it ("Cloud computing with mobile terminals") and in both cases failing to exploit its benefits. At the physical layer of current mobile technology implementations (LTE and UMTS), network layer multicast message is translated into a group of unicast messages (we will call this operation "unicast bundling"). Let BW(n) be the Bandwidth consumed by sending the same data to n hosts, then:

$$BW(n) = n \times BW(1) \qquad (2.1)$$

From network edge perspective, the transmitted message might be 100% efficient (i.e. sending a multicast message to n hosts), but from network core perspective and the system as a whole the efficiency is (1/n). The current physical layer implementation in mobile technologies does not suit applications using multicast/broadcast communications. We deduce that abstracting the mobile network as "Cloud computing with mobile terminals" proposed is definitely not a future proof solution. We believe that utilizing multicast is not a feature anymore, but a necessity to overcome the drawbacks generated from "unicast bundling".

Network edge solutions are not able to force the core network to forward multicast traffic without utilizing "unicast bundling". In order to solve this problem and make mobile cloud more agreeable and efficient, network core adaptations should be considered.

When designing our architecture, we seriously took into consideration users' and operators' concerns since we are completely aware that this architecture can make it to production only if it overcomes its competitors in performance and offers a valid business model that convinces investors and decision makers. We also designed this architecture to handle separately both types of traffic (unicast and multicast) to achieve elevated revenues and optimized utilization of resources. In the coming section we propose our architecture titled

OCMCA (Operator-Centric Mobile Cloud Architecture) as a response for the previously discussed concerns and a lesson learnt from previous architectures' mistakes.

## 2.4.1. Architecture Description

The lessons learnt from section 2.3, corresponds to users' concerns, are summarized as follows:

- **Delay**: "Cloud computing with mobile terminals" suits applications requiring complex processing because it is equipped with powerful cloud servers. Oppositely, "Virtual cloud computing provider" suits light applications because processing is executed close to the mobile device. Our architecture uses powerful cloud servers, but these servers should be closer to the mobile device.

- **Power consumption**: Similarly "Cloud computing with mobile terminals" suits applications requiring complex processing and "Virtual cloud computing provider" suits light applications. Our architecture offloads the job into powerful cloud servers thus will be suitable for complex applications. By decreasing delay as shown above, the power consumed by the mobile device waiting for the reply will be decreased.

- **Cost**: Wi-Fi traffic is cheaper than unicast mobile traffic. Our architecture uses multicast traffic when possible to decrease users' costs and increase operator's revenues as shown in the business model (section 2.4.4).

- **Privacy**: Executing a job in a trusted environment has higher privacy level. Our architecture positions the cloud servers in a trusted environment to ensure privacy.

- **Mobility**: The operator's network coverage and handover mechanisms make a connected user (connected to his services through operator's network) benefit from high mobility. Our architecture uses the mobile network to connect between cloud servers and users.

- **Scalability**: Statistical multiplexing proves that unified resources are more scalable, and this is why "Cloudlet" was not. Our architecture uses cloud servers accessible by any connected user (connected to the mobile operator's network) and connected with high bandwidth and low latency link to the operator's core network which makes it scalable. Its scalability can be increased through federation.

Figure 2.4 represents a normal EPS network with minimum entities needed for the basic functionalities (Call and Internet Access).

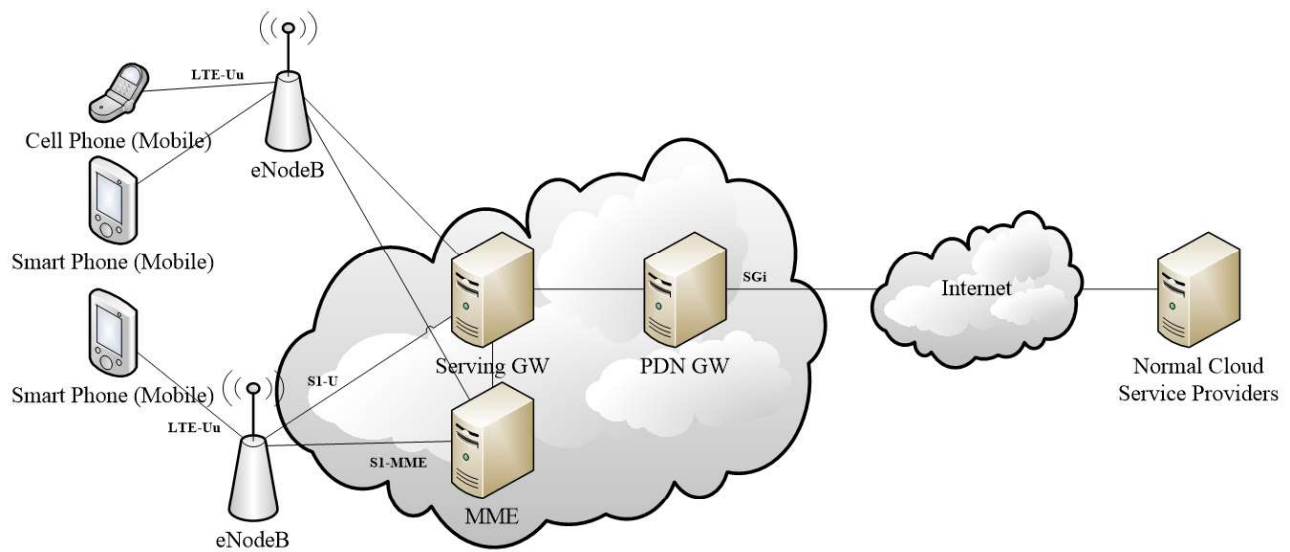

Figure 2.4 EPS network (minimum configuration)

Our architecture proposes the allocation of a cloud (called "Cloud server" throughout the remaining of this chapter) hosted by the mobile operator (in the mobile operator's premises or in a trusted network as shown in figure 2.5). "Cloud server's" location within the operator network is responsible for the following features:

- Close to the mobile core in order to eliminate the "delay" resulting from accessing the internet. The decreased distance can be seen in figure 2.5.
- Connected to the MME allowing it to differentiate between unicast and "multicast" traffic which facilitates separate handling of these two traffic types.
- Under the supervision of the mobile operator or a trusted party. The operator is also trusted by the user, since an SLA is already signed, thus the "Cloud server" is considered a "private" environment.

- The cloud is accessible by all connected users, since a direct connection exists with the core network. This feature is responsible for high user "mobility" and participates in increasing the "scalability" of this architecture.

The positioning of the "Cloud server" is graphically presented in figure 2.5.

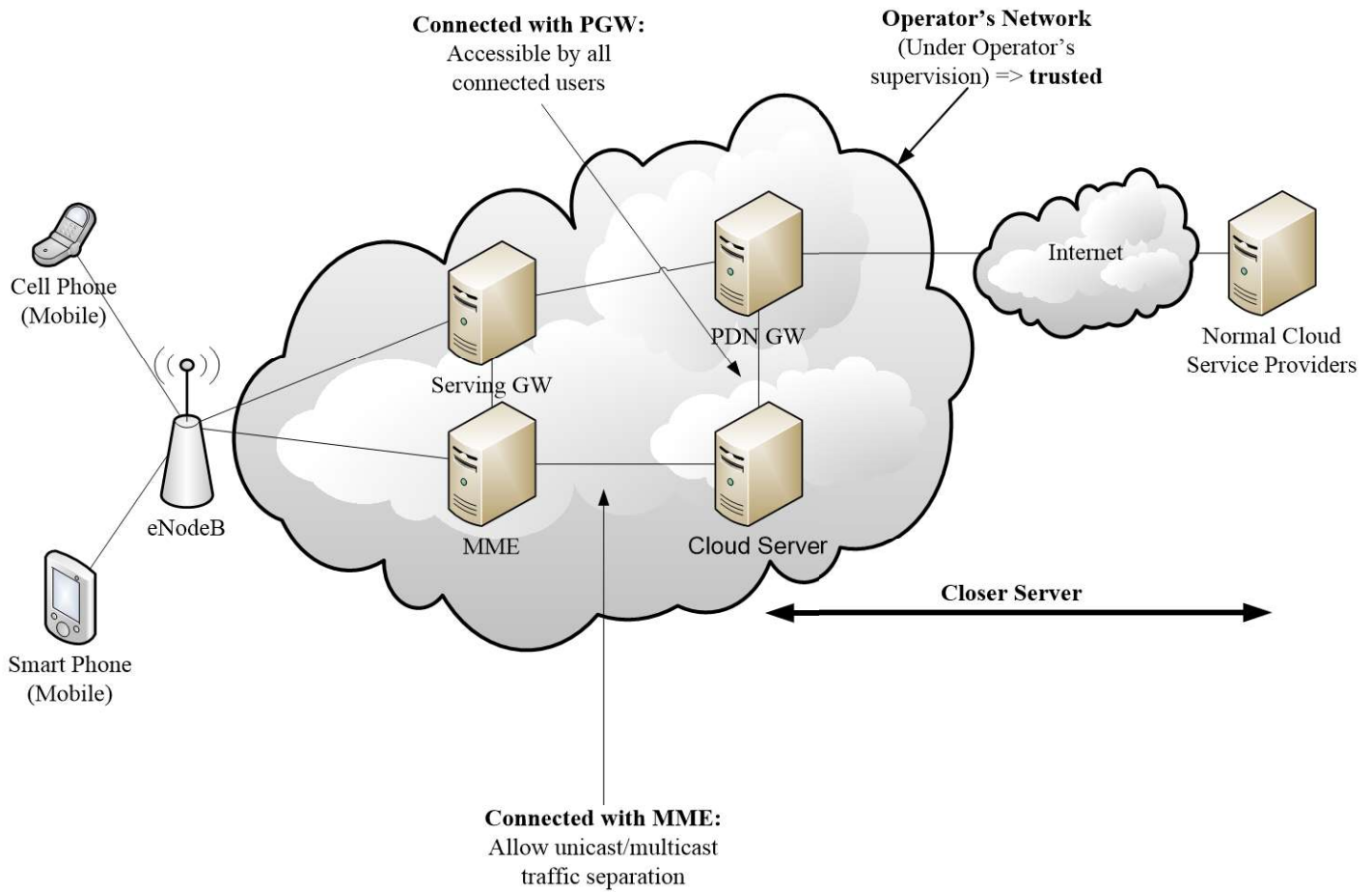

Figure 2.5 Initial phase of OCMCA (without multicast)

By positioning the "Cloud server" within the operator's network, we achieved all the features previously mentioned. Although the multicast traffic has been specified, it cannot be forwarded to the users without additional support. 3GPP has standardized multicast and broadcast packet transmission in UMTS and broadcast transmission in LTE through a feature named MBMS (Multimedia Multicast/Broadcast Service) [75][76][77][78] which was defined in 3GPP's technical specification as follows:

- "MBMS is a point-to-multipoint service in which data is transmitted from a single source entity to multiple recipients" [76].

- Physical broadcasting allows network resources to be shared when transmitting the same data to multiple recipients [76].

- Its architecture ensures efficient usage of radio-network and core-network resources, especially in radio interface [76].

MBMS requires the introduction of two nodes to the EPS's core (also named EPC: Evolved Packet Core) network [76], which are:

- **MBMS GW**: responsible for the connections with content owners, cloud servers in our case.

- **BM - SC (Broadcast Multicast Service Centre)**: provides a set of functions for MBMS User Services.

When the "Cloud server" receives a request it checks if the data should be unicast or multicast. In case of multicast, "Cloud server" notifies MME which in turn requests the P-GW through S-GW to add the participant to a multicast group. When the user is ready to receive the multicast traffic, the "Cloud server" forwards the data to BM-SC which in turn forwards it back to MBMS GW which is responsible for targeting the needed eNBs.

If the operator is providing multicast services (such as mobile TV), then MBMS GW and BM-SC are already installed and could be utilized by our architecture without the need for additional investment. Enabling multicast in LTE allows the local cloud, located within the mobile operator's premises, to transmit data efficiently to the users of one or more cells. Unicast based applications will use the normal "user dedicated channels" to send their unicast data to a specific user. The full configuration of our proposed architecture is presented in figure 2.6.

Cloud Server has been carefully positioned, in direct connection with P-GW using the standardized "SGi" interface, to require minimum standardization and development effort. Using this interface, makes the development and integration of "Cloud server" much easier ("SGi" libraries are already developed by vendors, no need for intermediate integration entities, etc.) in addition to the possibility to be co-located with the IMS (IP Multimedia Subsystem). In other words, IMS services can be installed as a service in the operator's cloud. This argument is also valid for other services offered by a mobile operator known as VAS (Value Added Services) such as: voice mail, top-up, credit transfer, waiting ringtone, etc.

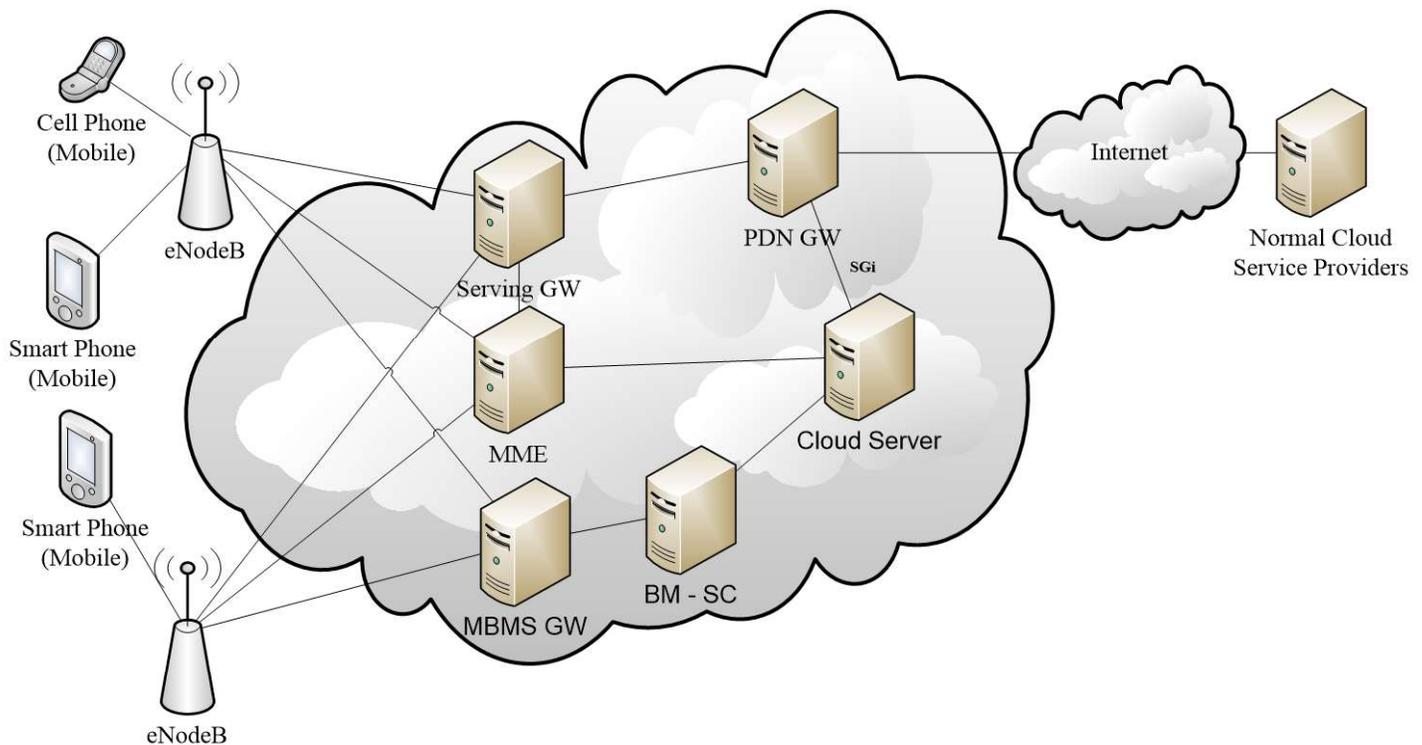

Figure 2.6 OCMCA (full configuration)

The added nodes (MBMS GW, BM-SC and Cloud Server) are transparent to normal mobile traffic. Mobile traffic will pass as follows:

- **Normal Mobile Traffic**: eNodeB, Serving GW, PDN GW (Packet Data Network GateWay). It is represented by the blue line in figure 2.7.
- **Unicast Cloud Traffic**: eNodeB, Serving GW, "Cloud Server", Serving GW, eNodeB. It is represented by the green line in figure 2.7.
- **Multicast Cloud Traffic**: eNodeB, Serving GW, "Cloud Server", BM-SC, MBMS GW. It is represented by the red line in figure 2.7.

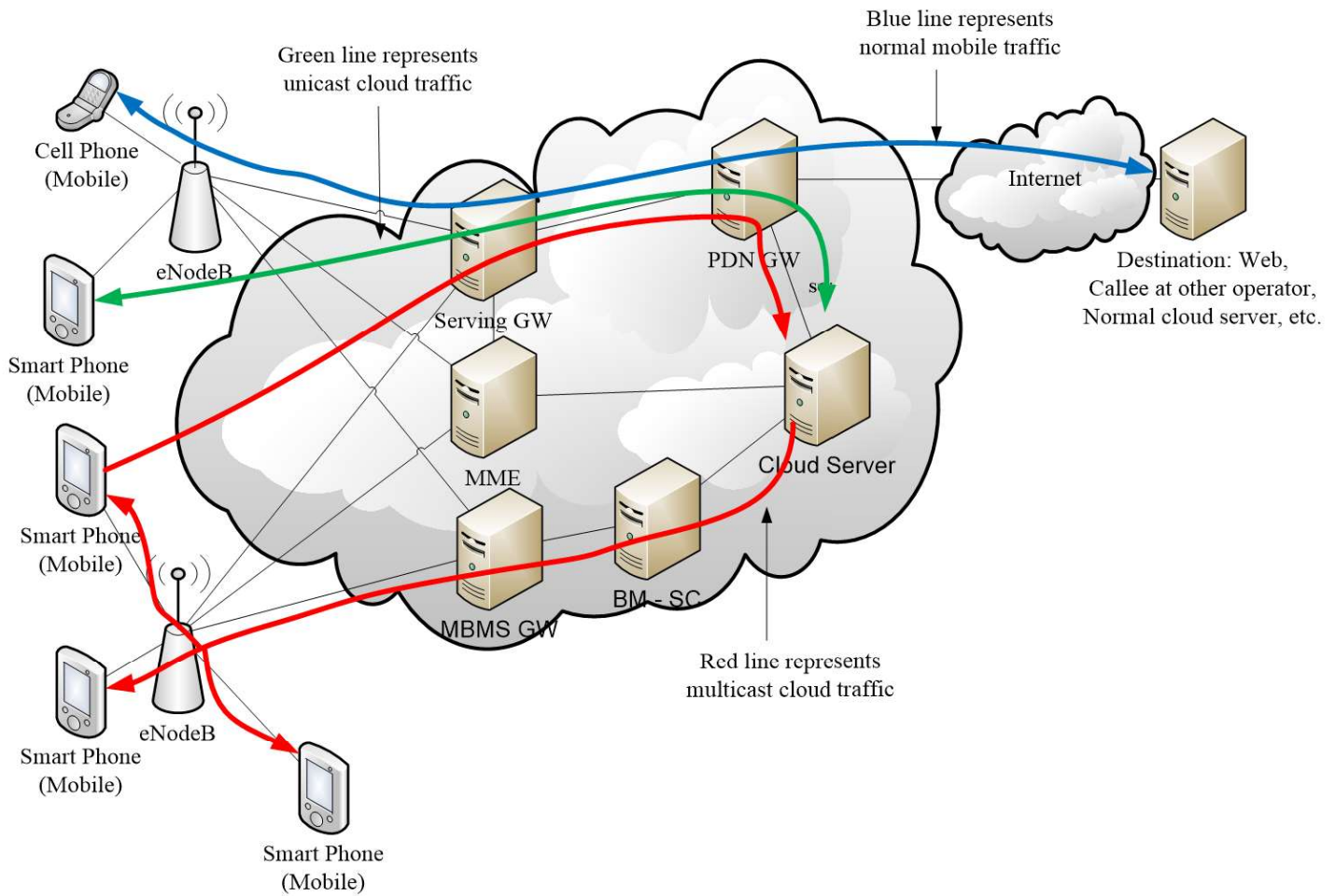

Figure 2.7 Traffic paths in OCMCA

## 2.4.2. User's Processing Environment

For a user to be able to execute her tasks at the "Cloud server" she has to be subscribed with either the mobile operator's cloud or any other cloud provider. In the first case, the user's environment is created and maintained at the "Cloud Server" so enjoying all the features mentioned in the previous section is possible. In the second case, the user's environment is located at the CSP which the user is subscribed to. The user environment has to be offloaded securely to the "Cloud server" which will be able then to offer its services in CSP's place. Offloading user environment is not possible without a trust relationship and financial motivation for the CSP. The most suitable method to establish monitored and accountable trust relationship between cloud providers is "cloud federation". Federation will be discussed in details in section 4.4.

Since users are usually connected with the same operator for long durations, we propose that the user's CSP should federate resources at the mobile operator's "Cloud server" and then offload the user's applications and environment settings. In this case the CSP will be able to offer its clients a faster and cheaper service while the mobile operator will be able to serve not just its clients, but any user subscribed with different cloud providers.

Enabling federation in OCMCA will lead to higher penetration rates since mobile operators don't need to enter in competition with mature cloud providers to gain market shares, while on the contrary cloud providers will be interested in creating a partnership with the mobile operator due to mutual profit opportunities. In this case, all computation-intensive processing is implemented within the mobile operator's network and the Terminal generated data are offloaded to the local cloud without the need to access the internet (which decreases the cost per bit of the transmitted data).

### 2.4.3. Performance evaluation

In this section, we present the simulation results of OCMCA using the same network configuration, application configuration and architecture requirements presented in section 2.3.3 in order to rank our architecture compared to the ones found in the literature. The simulation code is also developed using C# and the results are used to rank the performance by OCMCA. The results are shown in table 2.14.

Table 2.14 OCMCA's simulation results

| Appl. | Quantifiable Requirements (Lower values are better) | | | Non-quantifiable Requirements (Higher values are better) | | | |
|---|---|---|---|---|---|---|---|
| | Cost ($10^6$ financial units) | Delay (ms) | Power consumption ($10^6$ power units) | Privacy | Mobility | Scalability | Multicast Capable? |
| 1 | > 0.1001 > 1.001 | 3903 | 140.203 | High | High | High | Yes |
| 2 | > 0.1001 | 85 | 25.603 | High | High | High | Yes |

| | < 1.001 | | | | | | |
|---|---|---|---|---|---|---|---|
| **3 & 4** | > 10.0001 < 100.001 | 183 | 2409.013 | High | High | High | Yes |
| **5** | > 0.0002 < 0.002 | 39 | 1.336 | High | High | High | Yes |
| **6 & 10** | > 1.0001 < 10.001 | 101 | 242.023 | High | High | High | Yes |
| **7** | > 0.1001 < 1.001 | 85 | 25.603 | High | High | High | Yes |
| **8 & 9** | > 0.0002 < 0.002 | 1038 | 31.306 | High | High | High | Yes |
| **11** | > 0.2 < 2 | 2950 | 134.5 | High | High | High | Yes |
| **12 & 13** | > 0.0002 < 0.002 | 48 | 1.606 | High | High | High | Yes |
| **14** | > 0.1001 < 1.001 | 3903 | 140.203 | High | High | High | Yes |
| **15** | > 0.0011 < 0.011 | 164 | 5.293 | High | High | High | Yes |

We compare the performance of OCMCA against the previously presented architectures in table 2.15.

Table 2.15 OCMCA's ranking across different requirements

| Appl. | Cost | Delay | Power consumption | Scalability, Mobility and privacy (same as Cloud computing with mobile terminals) |
|---|---|---|---|---|
| **1** | 1 | 1 | 1 | 1 |
| **2** | 1 | 2 | 3 | 1 |
| **3 & 4** | 1 | 1 | 1 | 1 |
| **5** | 1 | 3 | 3 | 1 |
| **6 & 10** | 1 | 2 | 3 | 1 |
| **7** | 1 | 2 | 2 | 1 |
| **8 & 9** | 1 | 1 | 1 | 1 |
| **11** | 1 | 1 | 1 | 1 |
| **12 & 13** | 1 | 1 | 1 | 1 |
| **14** | 1 | 1 | 1 | 1 |
| **15** | 1 | 1 | 1 | 1 |

As can be seen in table 2.15, OCMCA achieves excellent scores in cost, mobility and privacy and good results in delay and power consumption in addition to centralized profits and fair fee distribution among users.

## 2.4.4. Business Model

We are going now to calculate the theoretical cost of application 7 using our proposed architecture and the three architectures discussed in section III. "Tour De France" is in its epic, racers are achieving close times and fans are surrounding the tracks of the race. The fans are viewing the race and keeping an eye on the ranking and timing of each racer, using a mobile cloud application. Let the cost of:

- 1 bit transmitted from a mobile phone to the "Serving GW" be noted as "A".
- 1 bit transmitted from the "Serving GW" to a mobile phone over a unicast channel be noted as "B".
- 1 bit transmitted from the "Serving GW" to a mobile phone over a multicast channel be noted as (B/m). Where m $\geq$1 since financial cost of multicast traffic is much cheaper than unicast.
- 1 bit transmitted from a mobile to another over a Wi-Fi interface be noted as "C".

Let the size of the request to get periodically the scores be "rq" bits, and the size of the retrieved data be "rp" bits sent every period T (we will consider the update rate to be once every minute). Let the average number of fans and riders within a Wi-Fi range using application 7 be N. Let the average number of fans and riders within a cell be M. The cost of application 7 over one hour period per user is shown in table 2.16.

In "Virtual cloud computing provider" architecture, one user will download the information from internet (downloader) and broadcast it to the remaining users (sharer) using Wi-Fi connection who in turn will resend the data in broadcast. Thus users are noted as "downloader" and "sharer".

Table 2.16 Costs of different architectures

| Architecture | Cost |
|---|---|
| Our proposed architecture | $rq \times A + 60 \times rq \times (B/m)$ |
| Cloud computing with mobile terminals | $rq \times A + 60 \times rq \times (B)$ |
| "Sharer" in Virtual cloud computing provider | $rq \times C + 60 \times rp \times C + N \times 60 \times rp \times C$ |
| "Downloader" in Virtual cloud computing provider | $rq \times A + 60 \times rp \times B + 60 \times rp \times C$ |
| Cloudlet | $rq \times C + 60 \times rp \times C$ |

$rq \times A$ is the cost to join a certain event. $60 \times rq \times (B/m)$ is the cost of receiving 60 updates in one hour over a multicast channel, considering that the scores are transmitted every minute. $60 \times rq \times (B)$ is the cost receiving 60 updates in one hour over a unicast channel. $60 \times rp \times C$ is the cost of broadcasting 60 updates to nearby users. $N \times 60 \times rp \times C$ is the cost of receiving broadcast updates from nearby users.

We are going now to compare the architectures based on financial cost. Wi-Fi is free (C=0), and multicast traffic is m times cheaper than unicast (where m is fixed by the operator). The cost will be in this case:

- Our proposed architecture: $60 \times (B/m) \times rq + rp \times A$

- "Cloud computing with mobile terminals" architecture: $60 \times B \times rq + rp \times A$

- "Cloudlet": free

- "Virtual cloud computing provider" architecture:

  o "Downloader": $60 \times B \times rq + rp \times A$

  o "Sharer": free

  o Average cost: $(60 \times B \times rq)/N + rp \times A$

$rp \times A$, rq and 60 are common to all users, so we are going to eliminate the previous parameters to better compare the costs. The simplified costs are shown in table 2.17.

Let us consider the number of participants to be 100 (M = N=100) and the cost of multicast traffic be 10 times cheaper (m=10). In this case the user pays for the studied application in OCMCA 10 times less than what he would pay for the same application in other architectures and the operator would receive (M/m) 10 times more profit than normal unicast for the same channel. Network availability is also preserved (since one channel is serving M users) in addition to the extra profits. This new business model creates the financial motivation for mobile operators to invest in mobile cloud computing and promises elevated revenues.

Table 2.17 Simplified costs of different architectures

| Architecture | User cost | Operator Revenues per channel | Comments |
|---|---|---|---|
| **Cloud computing with mobile terminals** | B | B | This architecture has good coverage and achieves good profits, but it is not scalable. The number of connected users is limited to "sub-carriers" available at this sector/cell. It also increases the customer rejection rate since the "subcarriers" will be already reserved. Higher customer rejection decreases potential profits and customer satisfaction. |
| **Virtual cloud computing provider** | Average: B/N Downloader: B Sharer: 0 | B | Although this architecture achieves low average user cost, but in reality only one user (Downloader) is paying full tariff and the remaining users (sharer) are getting advantage of it. The Downloader's motivation should be studied more for this architecture to implantable. |
| **Cloudlet** | 0 | 0 | Although this architecture seems very attractive for users, it creates both mobility and coverage issues; since most of "Tour De France's" race tracks are not covered by Wi-Fi signals. This architecture can only serve a small portion of the fans. |
| **OCMCA** | B/m | M×B/m | Usually M > m > 1, this architecture decreases the user cost and increases the operator's profit without affecting potential profits and decreasing customer satisfaction. This win-win situation is definitely interesting for operators. |

## 2.5. Chapter summary

We have shown in this chapter, that OCMCA achieves excellent scores in the majority of architecture requirements and a superior performance against all competitive architectures. It also achieves a business model that motivates investors due to the very promising revenues and limited investment needed (for instance: only software upgrade is needed to run OCMCA if IMS and MBMS servers are already installed at the operator's premises).

Since OCMCA can offer IMS as one of its SaaS services, other operator functionalities can be migrated to this cloud such as billing (also known as Intelligent Network), VAS, etc. This opens a huge door on the different models a mobile operator can be such as:

- **Outsourced Services**: The telecom vendor or any other CSP can offer VAS services to be rented by small operators that aren't able to deploy all the services in their network.
- **Shallow Operators**: An operator can outsource all its core services from a CSP and just maintain its radio network.
- **Virtual Operator**: An operator can outsource all its core services from a CSP and rent its radio network from other operators. This result in zero investment needed to join the mobile market.
- **No Operator**: A mobile network can be developed from HeNBs (Home eNBs) connected over the internet to a cloud-hosted core network which eliminates the need for fixed line operators.

The proposed operator models are all possible, but further research is needed to decide which model is capable of surviving till the deployment phase.